\documentclass[preprint2,tighten]{aastex6}
\pdfoutput=1 

\usepackage{amsmath,amstext}

\makeatletter
\def\env@matrix{\hskip -\arraycolsep 
  \let\@ifnextchar\new@ifnextchar
  \array{*{\c@MaxMatrixCols}c}}
\makeatother

\usepackage[T1]{fontenc}
\usepackage{apjfonts} 
\usepackage[figure,figure*]{hypcap}

\usepackage{booktabs}
\usepackage{threeparttable}
\usepackage{tabularx}

\hypersetup{linkcolor=black}
\usepackage{color, colortbl}
\definecolor{Gray}{gray}{0.95}

\usepackage{float}
\restylefloat{table}

\shortauthors{Sun et al.}

\begin{document}

\title{A Foreground Masking Strategy for [C\,{\small II}] Intensity Mapping Experiments Using Galaxies Selected by Stellar Mass and Redshift}

\author{G.~Sun\altaffilmark{1}}
\author{L.~Moncelsi\altaffilmark{1}}
\author{M.P.~Viero\altaffilmark{2,1}}
\author{M.B.~Silva\altaffilmark{3}}
\author{J.~Bock\altaffilmark{1,4}}
\author{C.M.~Bradford\altaffilmark{1,4}}
\author{T.-C.~Chang\altaffilmark{4,5}}
\author{Y.-T.~Cheng\altaffilmark{1}}
\author{A.R.~Cooray\altaffilmark{6,1}}
\author{A.~Crites\altaffilmark{1}}
\author{S.~Hailey-Dunsheath\altaffilmark{1}}
\author{B.~Uzgil\altaffilmark{7,1}}
\author{J.R.~Hunacek\altaffilmark{1}}
\author{M.~Zemcov\altaffilmark{8,3}}
\email{gsun@astro.caltech.edu}

\altaffiltext{1}{California Institute of Technology, 1200 E. California Blvd., Pasadena, CA 91125}
\altaffiltext{2}{Kavli Institute for Particle Astrophysics and Cosmology, Stanford University, 382 Via Pueblo Mall, Stanford, CA 94305}
\altaffiltext{3}{Kapteyn Astronomical Institute, University of Groningen, Landleven 12, NL-9747AD Groningen, the Netherlands}
\altaffiltext{4}{Jet Propulsion Laboratory, 4800 Oak Grove Drive, Pasadena, CA 91109}
\altaffiltext{5}{Institute of Astronomy and Astrophysics, Academia Sinica, Roosevelt Rd, Taipei 10617, Taiwan}
\altaffiltext{6}{Dept. of Physics \& Astronomy, University of California, Irvine, CA 92697}
\altaffiltext{7}{Max-Planck Institut f\"{u}r Astronomie, K\"{o}nigstuhl 17, D-69117, Heidelberg, Germany}
\altaffiltext{8}{Rochester Institute of Technology, Rochester, NY\ 14623}

\begin{abstract}
Intensity mapping provides a unique means to probe the epoch of reionization (EoR), when the neutral intergalactic medium was ionized by the energetic photons emitted from the first galaxies. The [C\,{\small II}] 158$\mu$m fine-structure line is typically one of the brightest emission lines of star-forming galaxies and thus a promising tracer of the global EoR star-formation activity. However, [C\,{\small II}] intensity maps at $6 \la z \la 8$ are contaminated by interloping CO rotational line emission ($3 \leq J_{\rm upp} \leq 6$) from lower-redshift galaxies. Here we present a strategy to remove the foreground contamination in upcoming [C\,{\small II}] intensity mapping experiments, guided by a model of CO emission from foreground galaxies. The model is based on empirical measurements of the mean and scatter of the total infrared luminosities of galaxies at $z < 3$ and with stellar masses $M_{*} > 10^{8}\,\rm M_{\rm \odot}$ selected in $K$-band from the COSMOS/UltraVISTA survey, which can be converted to CO line strengths. For a mock field of the Tomographic Ionized-carbon Mapping Experiment (TIME), we find that masking out the ``voxels'' (spectral-spatial elements) containing foreground galaxies identified using an optimized CO flux threshold results in a $z$-dependent criterion $m^{\rm AB}_{\rm K} \la 22$ (or $M_{*} \ga 10^{9} \,\rm M_{\rm \odot}$) at $z < 1$ and makes a [C\,{\small II}]/CO$_{\rm tot}$ power ratio of $\gtrsim 10$ at $k=0.1$ $h$/Mpc achievable, at the cost of a moderate $\la 8\%$ loss of total survey volume. 
\end{abstract}


\section[]{INTRODUCTION}

The formation of stars in the first generations of galaxies is closely associated with the Epoch of Reionization (EoR) occurring at $6 \la z \la 10$, during which Lyman continuum photons ionized the mostly neutral intergalactic medium (IGM) after recombination ($z\sim1100$). Advances in surveys of individual high-redshift galaxies at both near-infrared 
\citep[e.g.,][]{Ellis_2013, Bouwens_2015, Oesch_2015, Livermore_2017} and millimeter/sub-millimeter wavelengths \citep[e.g.,][]{Capak_2015, Carilli_2016}, together with constraints on the global ionization history from the cosmic microwave background \citep{Planck_2016tau} and a variety of spectroscopic diagnostics of the evolving IGM neutrality (see \citealt{R15} for a compilation), have greatly deepened our understanding of the reionization era over the past several years. However, none of these observables directly probes the entire ionizing photon budget responsible for reionization --- even for a typical ``ultra-deep'' survey with the most powerful telescopes like JWST, limitations on the sensitivity may result in missing up to 50\% of the total star formation inside galaxies at $z>8$, given the steep faint-end slope of the galaxy luminosity function implied by current observations \citep{SF_2016,Furlanetto2017}.

An alternative to galaxy counting is to measure the aggregate emission from all galaxies through line intensity mapping. In this approach, an imaging spectrometer is used to map the surface brightness of the Universe as a function of position on the sky and frequency. A bright emission line creates structure in the resulting 3D map due to the cosmic matter distribution; this structure is analyzed in the Fourier domain, i.e. with a power spectrum. In particular, the variance on large scales carries information about the total line emission from all galaxies, integrated over the full luminosity function, including all faint sources \citep{VL_2010,VTL_2011}. 

[C\,{\small II}] is a particularly promising probe for line intensity mapping of the reionization epoch \citep[e.g.,][]{Gong_2012, Silva_2015, Yue_2015, Breysse_2015, Serra_2016}. As the dominant coolant of the cold, neutral interstellar medium (ISM), the [C\,{\small II}] 157.7$\mu \rm{m}$ fine-structure line is among the strongest emission lines in aggregate galaxy spectra and it is found to be a reliable tracer of the star formation activity of typical star-forming galaxies \citep{DeLooze_2011, Herrera-Camus_2015}. Observationally, [C\,{\small II}] is redshifted into the 200--300\,GHz atmospheric window, which is relatively accessible from even modest millimeter-wave sites. 

However, extracting signals from EoR galaxy populations in intensity mapping experiments is challenging because these galaxies are typically not the dominant source of fluctuations in a map. EoR signals suffer from both a small luminosity density and the $1/D_L^2$ cosmological dimming relative to the later-time emission when luminosity density was at its peak. Specifically, for an intensity mapping experiment at $\sim250\,\rm{GHz}$, the EoR [C\,{\small II}] signal will be confused by the CO rotational lines emitted by foreground galaxies ($3 \leq J_{\rm upp} \leq 6$, at $0<z<2$) and redshifted into the same frequency band, in addition to the continuum sources that make up the cosmic infrared background (CIB). As a result, an accurate measurement of the EoR [C\,{\small II}] power spectrum requires that foreground contamination can either be appropriately identified and subtracted, or masked. 

A variety of foreground removal techniques for general line intensity mapping experiments have been proposed for continuum foregrounds and/or line interlopers. Treatments of continuum emission are especially well-studied for extracting the cosmological 21cm signal and often exploit spectral smoothness, which allows a suite of subtraction or avoidance techniques \citep[e.g.,][]{FOB_2006,McQuinn_2006,Harker_2009,LiuTegmark_2011,Parsons_2012,Chapman_2016}. As the continuum-to-line brightness in [C\,{\small II}] measurements is smaller by orders of magnitude, we expect these 21cm methods will prove effective.

Line interlopers, such as the CO signal in the [C\,{\small II}] EoR band, on the other hand, are different in that they are truly 3D signals. Therefore they require different cleaning techniques. One approach is cross-correlating the target line with an alternative tracer of the same cosmic volume such as galaxy surveys \citep{VL_2010,Gong_2012,Gong_2014,Silva_2015}. Another promising approach is ``line de-confusion'', introduced by \citet{VL_2010} and studied in detail recently by \citet{LT_2016} and \citet{Cheng_2016}, which uses the fact that the CO foreground power spectra projected onto the [C\,{\small II}] coordinate system are highly anisotropic between the directions perpendicular and parallel to the light of sight. 

In this paper we focus on what is arguably the simplest approach that works in real space: voxel masking. The masking approach consists of identifying foreground galaxies in 3D using external galaxy catalogs and removing the corresponding voxels from the survey. This ``guided'' masking approach is fundamentally different from the blind, bright-voxel masking approach discussed in \citet{Gong_2014} and \citet{Breysse_2015}, which works well only when the bright end of the voxel intensity distribution is dominated by the foreground, while all the signal is at the faint end (see Figure 9 of \citealt{Gong_2014}). However, while we expect that some of the foreground sources will be bright and directly detectable, faint sources likely contribute a large fraction of the CO foreground, based on the observed shape of CO luminosity function \citep[e.g.,][]{Walter_2014}. For example, the expected CO clustering signal at 250\,GHz may be 2--10 times larger than the [C\,{\small II}] signal, so up to 99\% of the integrated CO luminosity function needs to be masked out. This implies that a blind, bright-voxel masking approach will be insufficient, as found by \citet{Breysse_2015}, and therefore foreground sources must be traced and masked down to a greater depth to ensure a sufficient reduction.

The voxels containing CO-emitting sources must be identified a priori so that they can be masked from the [C\,{\small II}] survey. Using CO measurements directly is currently impractical because CO line surveys of individual galaxies are extremely time-consuming and may be feasible for only the brightest galaxies, while accurately measuring CO power spectra at intermediate redshifts is still an emerging field \citep[e.g.,][]{Walter_2014,Keating_2016,Decarli_2016}. We do note that some blind, deep CO surveys are underway with ALMA (PIs: Walter, Decarli), but even these do not scale to the cosmic volume (area and spectral range) required for the first-stage [C\,{\small II}] EoR intensity mapping experiments.

Alternatively, ancillary datasets (i.e. CO proxies) can be used to model both the position and brightness of foreground CO sources, in which case the masking depth required to sufficiently remove the foreground will depend on the uncertainty in the CO flux estimated with the proxy. A potential proxy for CO emission is the total infrared luminosity, believed to be proportional to star formation rate through the \citet{Kennicutt1998} relationship. Strong correlations are measured between the luminosities of various CO transitions and the total infrared luminosity for both local system and at $z\sim1\mbox{--}2$, albeit for relatively luminous galaxies \citep{CW_2013}. The limitation though comes from the lack of direct far-IR data to the required depth. For example, Spitzer MIPS serves as an excellent tracer of total infrared luminosity at $0.5<z<2$ \citep{Bavouzet_2008}. However, the source density required to sufficiently reduce the CO foreground, which we estimate to be $\sim 10^{5}\,\rm{deg^{-2}}$, is about twice as high as that of the deepest MIPS catalog. 

Fortunately, recent deep near-IR catalogs do have sufficient source density to potentially identify CO emitters down to the required depth. The challenge is to understand the degree to which the near-IR measurements can serve as a proxy for CO emission; this is the major thrust of this work. Our approach is to start from ultra-deep, near-infrared selected source catalogs and cross-correlate them with far-infrared/sub-millimeter maps  via stacking analysis to measure the mean infrared luminosities of galaxies \citep{Viero_2013, Schreiber_2015, Viero_2015} as well as the scatter in their population. The multi-wavelength coverage of these catalogs allows for high-quality photometric redshifts, which we use to position the foreground galaxies into our voxel space.

To estimate the CO foreground level  --- complete with mean and scatter --- and explore the effects that different levels of masking have on the resulting power spectrum, we first model the mean total infrared luminosity ($L_{\rm IR[8-1000\mu m]}$, or simply $L_{\rm IR}$ hereafter) as a function of stellar mass and redshift, and then exploit the empirical relationship between $L_{\rm IR}$ and $L^{\prime}_{\rm CO}$ to convert $L_{\rm IR}$ to CO luminosities, after including the scatters in both the $L_{\rm IR}(M_{*},z)$ and the $L_{\rm IR}$--$L^{\prime}_{\rm CO}$ correlation. Finally, as an application of our method, we use the CO power spectrum to determine the degree of masking necessary to significantly detect the [C\,{\small II}] power spectrum with the Tomographic Ionized-carbon Mapping Experiment (TIME, \citealt{Crites_2014}) at the angular scales of interest. It is important to note that our proxy-based method always allows for ``over-masking'', namely removing foreground galaxies that do not emit appreciable CO by discarding more voxels than is necessary, without biasing the EoR signal. This relies on the fact that the CO emission is uncorrelated with the target [C\,{\small II}] emission from the masked voxels, and that effects of masking such as mode mixing can be appropriately corrected \citep[e.g.,][]{Zemcov_2014}.

This paper is arranged as follows. In Section 2, we model the mean total infrared luminosity of galaxies as a function of stellar mass and redshift with the simultaneous stacking formalism and algorithm developed by \citet[][{\footnotesize SIMSTACK}\footnote{\href{https://web.stanford.edu/~viero/downloads.html}{https://web.stanford.edu/$\sim$viero/downloads.html}}]{Viero_2013}. We also describe in detail the innovative technique of thumbnail stacking on residual maps, used to characterize the scatter in $L_{\rm{IR}}$. We discuss the observational implications for the masking strategy of [C\,{\small II}] intensity mapping experiments in Section 3 and briefly conclude in Section 4. Throughout this paper, we assume a Chabrier \citeyearpar{Chabrier2003} initial mass function (IMF) and a flat, $\Lambda$CDM cosmology consistent with the most recent measurement by the \citet{Planck2015XIII}.

\section[]{METHODS FOR MODELING INFRARED GALAXIES AS CO PROXIES}

We model both the mean and variance of the galaxy total infrared luminosity in galaxy samples binned in redshift and stellar mass. We measure these quantities using an extension of the {\footnotesize SIMSTACK} method introduced by \citet{Viero_2013}. The modeled $L_{\rm IR}$ can then be related to the strength of CO emission from foreground galaxies. The results presented in this work are performed on the COSMOS field \citep{Scoville2007} by combining a catalog derived using the imaging described in \citet{Laigle_2016} but processed by the \citet{muzzin2013} pipeline, with maps spanning the full far-infrared/sub-millimeter (FIR/sub-mm) spectral range of the thermal spectral energy distribution (SED) from interstellar dust. Note that, in addition to the maps used in \citet{Viero_2013}, we use maps at $450 \mu$m and $850 \mu$m from deep SCUBA-2 observations made available by \citet{Casey_2013}, which provide critical constraints on the low-energy end of the SED \citep[for details on the fitting routine, see][]{Moncelsi2011,Viero2012}. The full dataset including the maps and catalog used is summarized in Table~\ref{tb:data} (see also \citealt{Laigle_2016}) and will be described in detail in Viero et al.\@ (in prep.). 

\subsection{Estimating the Mean $L_{\rm IR}(M_{*},z)$ with {\footnotesize SIMSTACK}}

{\footnotesize SIMSTACK} is an algorithm that takes galaxy positions from an external catalog, splits them into subsets (typically, but not necessarily, by stellar mass and redshift), and generates mock map layers that are \emph{simultaneously} regressed with the real sky map to estimate the mean flux density of each subset. Formally, it is an extension of simple thumbnail stacking \citep{Marsden_2009}, the difference being that the off-diagonal entries in the subsets covariance matrix are not assumed to be zero, so as to account for galaxy clustering. The simultaneous fitting provides a solution to the limitations of stacking in highly confused maps (i.e., biased flux density estimates due to the clustering of sources at angular scales comparable to that of the FIR/sub-mm beam), such that in the theoretical limit where the catalog is complete it naturally leads to a completely unbiased estimator (see Appendix~\ref{sec:A1} for some justification). \citet{Viero_2013} show that {\footnotesize SIMSTACK} yields unbiased results at any beam size, while conventional thumbnail stacking (e.g., ``median'' or ``mode'' stacking, etc.), without additional corrections, inevitably leads to wavelength-dependent biases, in the presence of galaxy clustering.

\begin{figure}[h!]
\includegraphics[width=0.5\textwidth]{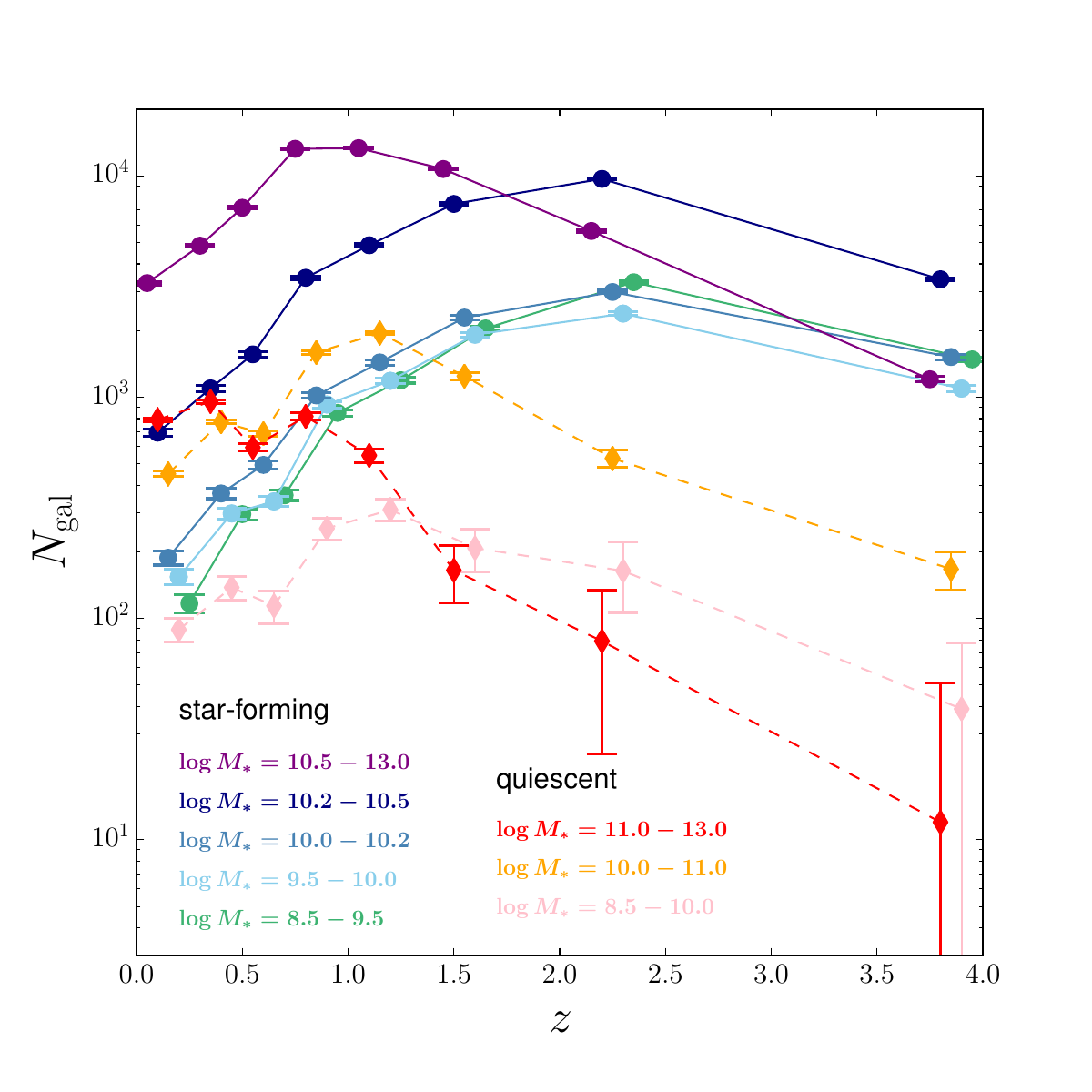}
\caption{
Numbers of star-forming or quiescent galaxies in bins of stellar mass and redshift. The binning is optimized to have more than 100 (10) star-forming (quiescent) galaxies in each bin and be approximately uniform in lookback time. The error bars show the square roots of the numbers of galaxies, which are Poisson distributed.
}
\label{fig:binning}
\end{figure}

The first step in measuring $L_{\rm IR}(M_{*},z)$ is to split the catalog into subsets of star-forming and quiescent galaxies based on their $U-V$ vs.\@ $V-J$ colors \citep[UVJ, e.g.,][]{Williams_2009}, and then again into bins of stellar mass (5 and 3 layers for star-forming and quiescent galaxies, respectively) and redshift (8 layers), determined by their optical and near-infrared photometry. We developed an algorithm to calculate the optimized locations of the $5\times8+3\times8=64$ stellar-mass/redshift bins so that each bin contains at least 100 (10) star-forming (quiescent) galaxies, as illustrated in Figure~\ref{fig:binning}.

\begin{table}[t!]
  \footnotesize
  \centering
  \begin{threeparttable}
  \begin{tabularx}{0.5\textwidth}{lcc}
    \toprule
    \toprule
    \multicolumn{3}{c}{\small \textbf{MAPS}} \\
	\midrule
    \textbf{Instrument/Telescope} & \textbf{Wavelength} & \textbf{1-$\sigma$ sensitivity}\tnote{a} \\
     & [$\mu$m] & [mJy/beam] \\
     &  & literature (measured) \\
    \midrule
    MIPS/\textit{Spitzer} & 24 & 0.06\tnote{b}\ \ (0.08)\\
    					  & 70 & 1.7\tnote{c}\ \ (2.85) \\
    PACS/\textit{Herschel} & 100 & 5\tnote{d}\ \ (3.1) \\
    					   & 160 & 10\tnote{d}\ \ (7.4) \\
    SPIRE/\textit{Herschel} & 250 & $\dagger$5.8\tnote{e}\ \ (6.8) \\
    						& 350 & $\dagger$6.3\tnote{e}\ \ (7.4) \\
    						& 500 & $\dagger$6.8\tnote{e}\ \ (7.7) \\
    SCUBA--2/\textit{JCMT} & 450 & $\dagger$4.7\tnote{f}\ \ (4.5) \\
    					   & 850 & $\dagger$0.8\tnote{f}\ \ (1.5) \\
    AzTEC/\textit{JCMT} & 1100 &  $\dagger$1.3\tnote{g}\ \ (1.6)\\
    \midrule
    \multicolumn{3}{c}{\small \textbf{CATALOG} (COSMOS/UVISTA DR2)} \\
	\midrule
    \textbf{Instrument} & \textbf{Filter} & \textbf{3-$\sigma$\,depth}\tnote{h} \\
    \textbf{/Telescope} & \textbf{/Central $\lambda$}\,[\AA] & \textbf{$\pm0.1$} \\
    \midrule
    \textit{GALEX} & NUV\,/\,$2.3139\times10^3$ & 25.5 \\
    MegaCam/CFHT & $u^*$\,/\,$3.8233\times10^3$ & 26.6 \\
    Suprime-Cam/Subaru & $B$\,/\,$4.4583\times10^3$ & 27.0 \\
    & $V$\,/\,$5.4778\times10^3$ & 26.2 \\
    & $r$\,/\,$6.2887\times10^3$ & 26.5 \\
    & $i^{+}$\,/\,$7.6839\times10^3$ & 26.2 \\
    & $z^{++}$\,/\,$9.1057\times10^3$ & 25.9 \\
    & $IA427$\,/\,$4.2634\times10^3$ & 25.9 \\
    & $IA464$\,/\,$4.6351\times10^3$ & 25.9 \\
    & $IA484$\,/\,$4.8492\times10^3$ & 25.9 \\
    & $IA505$\,/\,$5.0625\times10^3$ & 25.7 \\
    & $IA527$\,/\,$5.2611\times10^3$ & 26.1 \\
    & $IA574$\,/\,$5.7648\times10^3$ & 25.5 \\
    & $IA624$\,/\,$6.2331\times10^3$ & 25.9 \\
    & $IA679$\,/\,$6.7811\times10^3$ & 25.4 \\
    & $IA709$\,/\,$7.0736\times10^3$ & 25.7 \\
    & $IA738$\,/\,$7.3616\times10^3$ & 25.6 \\
    & $IA767$\,/\,$7.6849\times10^3$ & 25.3 \\
    & $IA827$\,/\,$8.2445\times10^3$ & 25.2 \\
    & $NB711$\,/\,$7.1199\times10^3$ & 25.1 \\
    & $NB816$\,/\,$8.1494\times10^3$ & 25.2 \\
    VIRCAM/VISTA & $Y$\,/\,$1.0214\times10^4$ & 25.3 \\
    & $J$\,/\,$1.2535\times10^4$ & 24.9 \\
    & $H$\,/\,$1.6453\times10^4$ & 24.6 \\
    & $K$\,/\,$2.1540\times10^4$ & 24.7 \\
    IRAC/\textit{Spitzer} & ch1\,/\,$3.5634\times10^4$ & 25.5 \\
    & ch2\,/\,$4.5110\times10^4$ & 25.5 \\
    & ch3\,/\,$5.7593\times10^4$ & 23.0 \\
    & ch4\,/\,$7.9595\times10^4$ & 22.9 \\
    \bottomrule
    \bottomrule
  \end{tabularx}
  \begin{tablenotes}
\item [a] {\footnotesize Dagger sign means the sensitivity is confusion-limited. The values in parentheses are estimated directly from the maps we used. }
\item [b] {\footnotesize \citet{Sanders_2007} }
\item [c] {\footnotesize \citet{Frayer_2009} }
\item [d] {\footnotesize Table 3.1, 
\url{http://herschel.esac.esa.int/Docs/Herschel/html/ch03s02.html}}
\item [e] {\footnotesize \citet{Nguyen2010} }
\item [f] {\footnotesize \citet{Chen_2013} }
\item [g] {\footnotesize \citet{Scott_2008} }
\item [h] {\footnotesize Limiting magnitudes are calculated from variance map in 2" aperture on PSF-matched images.}

  \end{tablenotes}
  \end{threeparttable}
  \caption{Map and catalog information. \label{tb:data}}
\end{table}

\begin{figure}[h!]
 \includegraphics[width=0.5\textwidth]{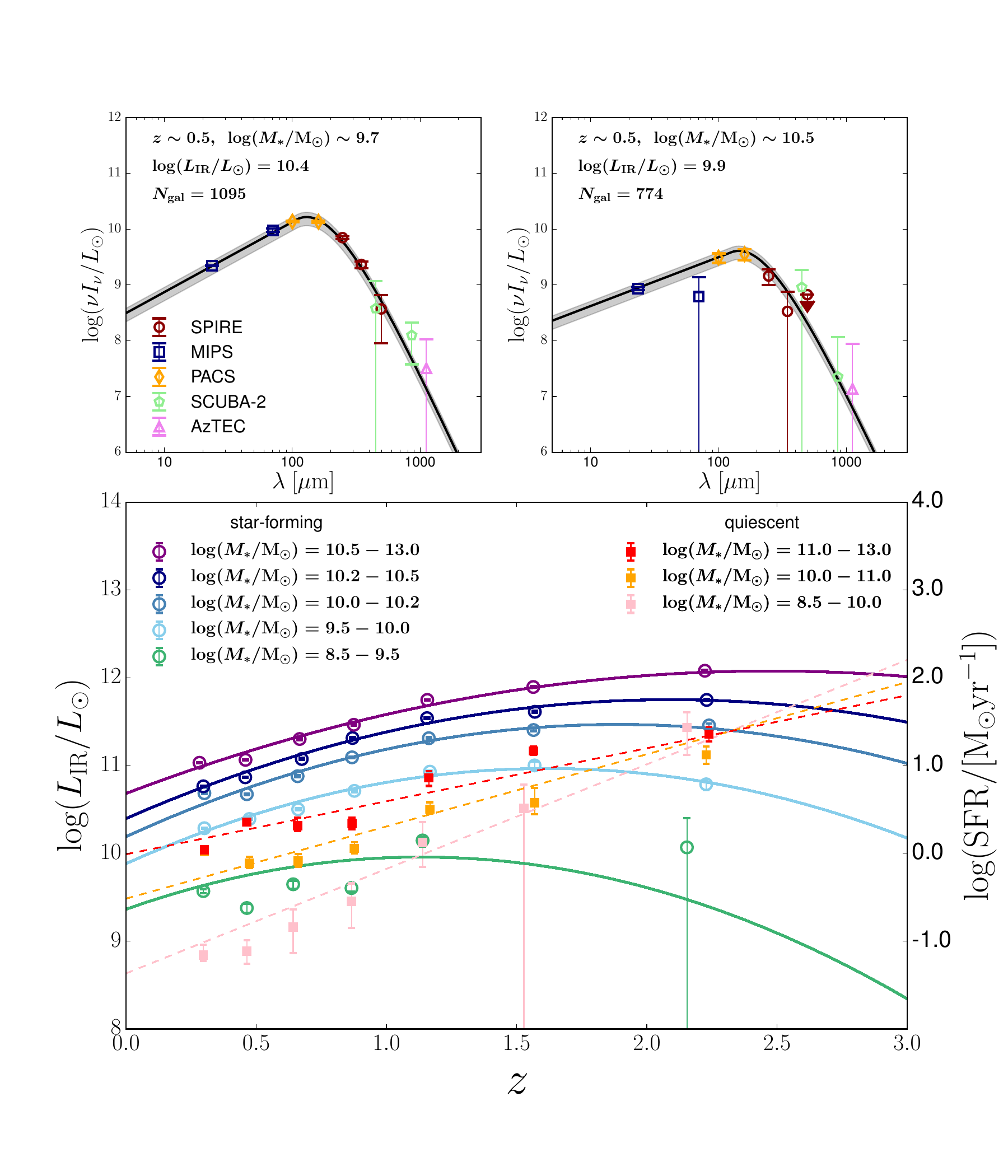}
 \caption{\textbf{Top}: Sample best-fit SEDs of star-forming (left) and quiescent galaxies (right). \textbf{Bottom}: Polynomial fits to the mean $L_{\rm IR}(M_{*},z)$ estimated from the stacked fluxes and best-fit, modified graybody spectra. Open (filled) markers show the measured luminosities in individual $(M_{*},z)$ bins for star-forming (quiescent) galaxies, while the solid (dashed) curves represent the corresponding best-fit curves.}
 \label{fig:LMZ}
\end{figure}

Next, the average FIR/sub-mm flux density for each bin and at each wavelength is estimated with {\footnotesize SIMSTACK}. Uncertainties on the mean flux densities are estimated with an extended bootstrap technique which takes into account the uncertainties in the photometric redshift and stellar-mass estimates of individual sources. $L_{\rm IR}$ for each bin is estimated by first fitting a modified blackbody (or graybody) with emissivity index $\beta = 2$, and the Wien side approximated as a power-law with slope $\alpha = -2$ \citep{blain2002}, to the full spectrum of intensities $\nu I_{\nu}(\lambda)$, and then integrating under the best-fit graybody from $\lambda_{\rm{rf}} = 8$ to $1000\, \mu$m. The final step is to fit the full set of mean $L_{\rm IR}$'s with multiple linear regression as described by \citet{Viero_2013}:
\begin{equation}
\log L_{\rm{IR}}(M_{*},z) = \sum_{p=0}^{n} \left\{\left[\sum_{q=0}^{n} A_{p,q} \left(\log M_{*}\right)^{q}\right] z^{p}\right\}, 
\label{eq:lmz_e1}
\end{equation}
where $n=2$ and 1 for star-forming and quiescent galaxies, respectively. The coefficient matrices $A_{p,q}$ are found to be
\begin{equation}
A_{p,q}^{\rm{sf}} = 
\begin{pmatrix} 
2.417 & 0.733 & 0.004 \\
-38.84 & 8.080 & -0.406 \\
4.947 & -1.223 & -0.069 
\end{pmatrix}
\end{equation}
and 
\begin{equation}
 A_{p,q}^{\rm{qt}} = 
 \begin{pmatrix}
  0.845 & 0.820 \\
  4.556 & -0.354
 \end{pmatrix}.
\end{equation}
Figure~\ref{fig:LMZ} shows two sample SED fittings to the stacked fluxes, together with the best-fit polynomials to the mean $L_{\rm IR}(M_{*},z)$ relations of star-forming and quiescent galaxies separately. As demonstrated in \citet{Viero2012}, the modified blackbody approximation produces mean SEDs consistent with best-fit templates such as \citet{CE01} and the derived mean $L_{\rm IR}$ is largely insensitive to the exact choice of the Wien side slope $\alpha$. 

\begin{table*}[t!]
\footnotesize
\centering
\begin{tabular}{  p{2cm}  p{11cm}  p{2.5cm}  }
\toprule
\toprule
\textbf{TERM} & \textbf{DESCRIPTION} & \textbf{REFERENCE} \\ [0.5ex] \hline \\ [-2ex]
 
        \textbf{Bins} & Stellar mass or redshift intervals used to divide galaxies into sub-populations for stacking analysis.  & S2.1  \\ \\ [-2ex]
        \textbf{Layer} & A subset of a real/mock sky image (or map) attributed to only the sources in the corresponding stellar mass or redshift bin. & S2.1, S2.2  \\ \\ [-2ex]
        \textbf{Scatter} & In this paper, we exclusively define ``scatter'' as the standard deviation of flux density or luminosity in the source population, which is characterized and represented by $\sigma_{S}$ in equation~\ref{eq:log_normal}. & S2, App.A, Eq.~\ref{eq:log_normal} \\ \\ [-2ex]
        \textbf{(Un)perturbed} & Fluxes being assigned to the sources in a specific layer are drawn from a distribution with the mean equal to the best-fit value given by {\scriptsize SIMSTACK} and some (zero) nonzero width defined by the scatter. & S2.2, App.A, Eq.~\ref{eq:log_normal}  \\ \\ [-2ex]
        \textbf{Real/Mock} & ``Real'' refers to the actual sky image, whereas ``mock'' refers to the image reconstructed using source locations and perturbed mean fluxes from {\scriptsize SIMSTACK}. More specifically, in our analysis we construct the mock sky image by merging 1) a layer of interest perturbed according to a distribution with a tunable scatter and 2) background layers perturbed by a distribution with a fiducial scatter of 0.3~dex. & S2.2, Eq.~\ref{eq:D_real},~\ref{eq:D_mock}  \\ \\ [-2ex]
        \textbf{Base} & The ``base'' map, different from the mock image, is obtained by merging 1) an unperturbed layer of interest and 2) background layers perturbed by a distribution with a fiducial scatter of 0.3~dex. & S2.2, Eq.~\ref{eq:D_real},~\ref{eq:D_mock}  \\ \\ [-2ex]
        \textbf{Residual} & The difference between the real or noise-added mock sky image and a ``base'' one. & S2.2, Eq.~\ref{eq:D_real},~\ref{eq:D_mock}  \\ \\ [-2ex]
        \textbf{$\mathbf{D}_{\rm{real}}$, $\mathbf{D}_{\rm{mock}}$} & A small cutout image a few pixels by side, where each pixel measures the standard deviation of a data cube obtained by thumbnail-stacking the residual map at the positions of the sources in each $i,\ j$ layer. & S2.2, Eq.~\ref{eq:D_real},~\ref{eq:D_mock}  \\ \\ [-2ex]
\bottomrule
\bottomrule
\end{tabular}
\caption{A summary of the terms used in our discussion of methodology in Section 2 and 3. }
\label{table:1}
\end{table*}

\subsection{Characterization of the Scatter in $L_{\rm IR}(M_{*},z)$} \label{calibration_form}

At this point, we have modeled the mean infrared luminosity as a function of stellar mass and redshift, but naturally we expect $L_{\rm IR}$ of individual galaxies to depart from this model, with some characteristic scatter.  The question we aim to answer now is 
what is the degree of scatter of the full ensemble of sources? 

The answer lies in the standard deviation of the \emph{residual map} (see Table~\ref{table:1} for the definition), which is the difference between the real sky map at each wavelength and a synthetic map made by applying the $L_{\rm IR}(M_{*},z)$ model to the original catalog (i.e., the actual stellar masses, redshifts, and sky positions). In a universe where (i) objects are perfectly described by the mean model with no scatter, (ii) catalogs are 100\% complete, and (iii) maps have no noise, the residual map would be completely blank. In practice, the actual residual map will have structure due to the intrinsic stochasticity of the galaxy populations, catalog incompleteness, as well as instrumental noise.

\begin{figure}[h!]
\includegraphics[width=0.5\textwidth]{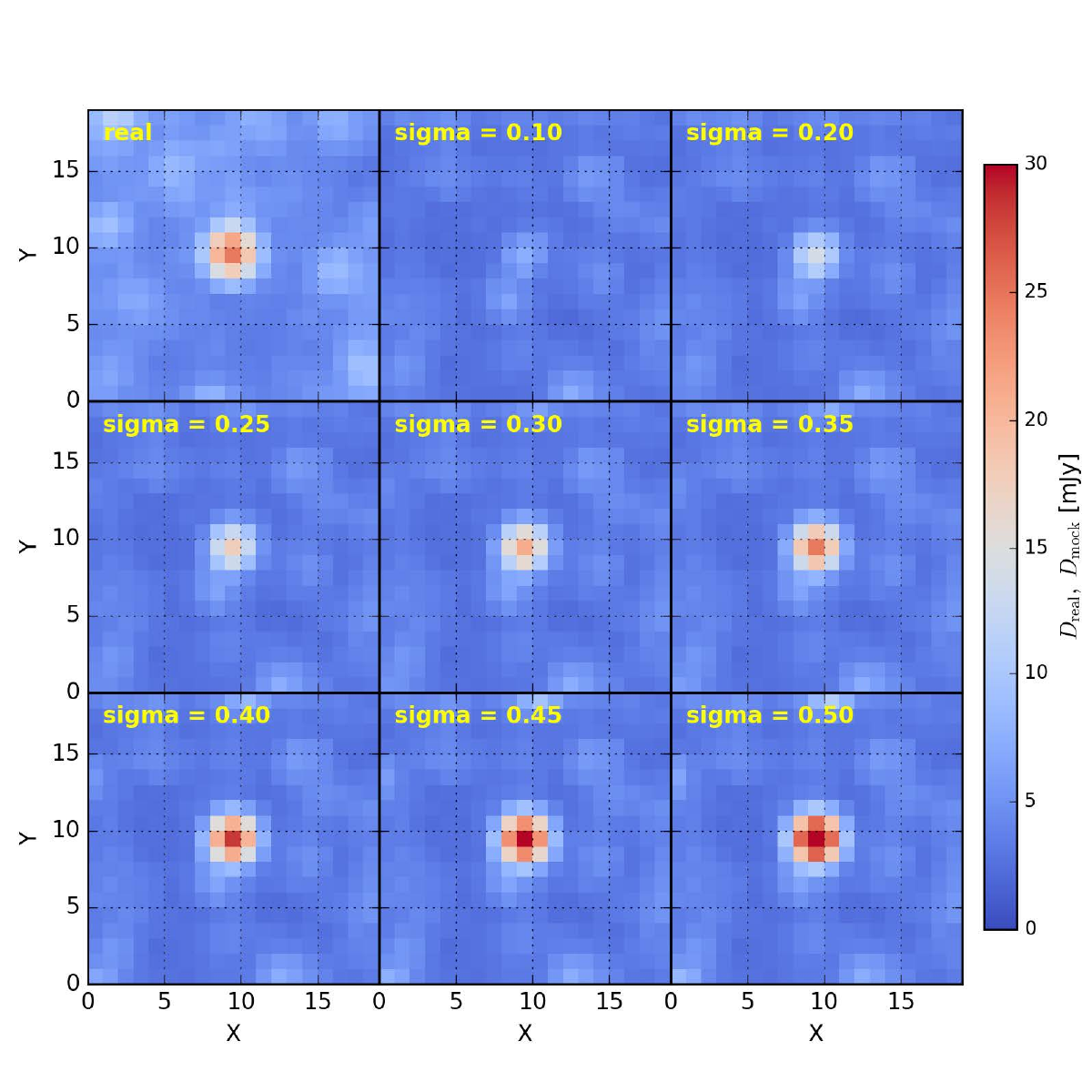}
\caption{
Standard deviation in thumbnail stacks, illustrating the scatter characterization method. The top left panel shows the standard deviation in the real residual (real map minus ``base'' using the mean relation). The other panels show mock residuals with various levels of log scatter (per equation~\ref{eq:log_normal}) artificially incorporated. This figure refers to a single bin: $0<z<0.3$, $\log M_{*}=10.5$--13. The central pixels show the standard deviation due to source variance -- a value of $\sigma_{S} \sim 0.35$ best reproduces the measured variance in the map.
}
\label{fig:map_illustration}
\end{figure}
 
We now introduce a method to formally characterize the scatter about the mean $L_{\rm  IR}(M_{*},z)$ relation by leveraging the structure in the residual map. Although our method has similarities with the \lq\lq scatter stacking\rq\rq\ method described in \citet{Schreiber_2015}, our use of residual maps ---  estimated by taking the difference with ``base'' maps generated with {\footnotesize SIMSTACK}-derived luminosities --- makes us less susceptible to clustering contamination, and provides a more robust estimate of the scatter in each $(M_{*},z)$ bin, validated through an extensive set of end-to-end simulations (see Section~\ref{sec:A1}). Due to the layered structure of our maps, the interplay between individual layers (often with a root-mean-square, or rms, amplitude below the confusion limit of the real map) must be investigated through simulations to estimate the scatter in each layer. For simplicity, we assume that the scatter is dominated by the stochasticity of the star-formation activity and therefore is independent of wavelength. We perform the scatter calibration with the 250$\mu$m SPIRE/HerMES \citep{griffin2010,oliver2012} map which covers the entire COSMOS/UltraVISTA field.

We assume that for a given redshift $z_{i}$ and stellar mass $M_{*,j}$, the actual flux density $S$ (and therefore the total infrared luminosity) is log-normally distributed about the mean value with a scatter $\sigma_{S}$, an assumption that is motivated by the observed scatter in the star-formation main sequence (SFMS, e.g., \citealt{Sargent_2012}). Namely,
\begin{equation}
	P \left(x|\mu, \sigma_{S}\right) = \frac{1}{\sigma_{S} \sqrt{2\pi}} \exp \left(-\frac{(x-\mu)^{2}}{2\sigma_{S}^{2}}\right), 
	\label{eq:log_normal}
\end{equation}
where $\mu = \log \langle S (M_{*},z) \rangle$ is the log of the mean flux density measured by {\footnotesize SIMSTACK}. Intuitively, $\sigma_{S}$ can be estimated by examining the statistics of source fluxes (e.g., standard deviation) in each stellar-mass/redshift bin of interest. However, for the highly-confused far-IR maps we use, clustering could render measured statistics biased by the contribution from sources in other bins, whose scatter must also be properly accounted for. 

Therefore we assign a fiducial scatter to the ``background'' sources. As will be shown in Section 3.2, the actual scatter can be measured without bias using our method, as long as it is not drastically different from the fiducial value. In particular, the scatter being investigated here is analogous to that of the SFMS, which can be explained as an application of the central limit theorem \citep{Kelson_2014} and is measured to be around $\sim 0.3$\,dex \citep{Behroozi_2013,Sparre_2015}. We therefore adopt 0.3\,dex as the fiducial population scatter for the ``background'' sources in our mock maps and demonstrate in Section 3 that it is indeed a reasonable choice. 

We hereafter refer to the actual sky image as the ``real map'' and the synthetic map based on the $L_{\rm{IR}}$ model as the ``mock map''. In addition, we call a layer unperturbed when the flux density of its sources is constant and equal to the average value $\mu$ found using {\footnotesize SIMSTACK}, while a layer is perturbed when each source has been assigned a flux density according to a log-normal distribution of mean $\mu$ and scatter $\sigma_{S}$. Finally, as anticipated before, the residual map is either a real or mock map from which the ``base'' map (the layer of interest, unperturbed, plus the background layers perturbed with the fiducial scatter) is subtracted. This nomenclature is summarized in Table~\ref{table:1}.

The crucial step of the algorithm is that we take the standard deviation $D_{\mathrm{real}}^{k}$, computed over the positions of all cataloged sources in the $(M_{*},z)$ bin of interest, of the residual real map, and compare it to its counterpart $D_{\mathrm{mock}}^{k}$, which is the standard deviation of a residual mock map obtained by adding up all perturbed layers, plus noise floor \citep[e.g.,][]{Nguyen2010}.

Mathematically, at a given pixel of interest $k$, we have
\begin{align}
D_{\mathrm{real}}^{k} &= \mathrm{SD} \biggl[ S_{\mathrm{real}}^{k} - \sum_{i,j} S_{\mathrm{base}}^{k}(z_{i}, M_{*, j}) \biggr]
\label{eq:D_real}
\end{align}
and
\begin{align}
D_{\mathrm{mock}}^{k} = &\ \mathrm{SD} \biggl[ \sum_{i,j} S_{\mathrm{mock}}^{k}(z_{i}, M_{*, j}) + S_{\mathrm{noise}}^{k} + \nonumber \\
						 &-\sum_{i,j} S_{\mathrm{base}}^{k}(z_{i}, M_{*, j}) \biggr]. 
\label{eq:D_mock}
\end{align}
where SD stands for taking the standard deviation of the thumbnail-stacked cube at each pixel $k$, and $i,\,j$ are the indices of redshift and stellar-mass bins (see Table~\ref{table:1} for a reminder of the definitions).

\begingroup
\renewcommand{\arraystretch}{1.3}
\newcolumntype{g}{>{\columncolor{Gray}}c}
\begin{table*}[t!]
\footnotesize
\begin{center}
\begin{threeparttable}
\begin{tabular}{c g c g c}  
\toprule
\toprule
\multicolumn{1}{c}{} & \multicolumn{4}{c}{\textbf{\boldmath{Number of Galaxies ($N_{\rm{gal}}$), Luminosity ($\log [L_{\rm{IR}}/L_{\odot}]$), and Scatter ($\sigma_{\rm L}\,[\rm{dex}]$)}}} \\
\cmidrule(r){2-5}
  & \boldmath{$0<z<0.3$} & \boldmath{$0.3<z<0.5$} & \boldmath{$0.5<z<0.7$} & \boldmath{$0.7<z<1$\tnote{a}} \\
\midrule
\multicolumn{5}{c}{\textit{Star-forming Galaxies}} \\
\cmidrule(l{18em}r{18em}){1-5} 
\boldmath{$10^{10.5} < \frac{M_{*}}{M_{\odot}} < 10^{13}$}       &  117, $10.94^{+0.01}_{-0.01}$, 0.33  &  296, $11.04^{+0.01}_{-0.01}$, 0.34    &  360, $11.25^{+0.01}_{-0.01}$, 0.35    &  849, $11.42^{+0.01}_{-0.01}$, 0.33      \\ 
\boldmath{$10^{10.2} < \frac{M_{*}}{M_{\odot}} < 10^{10.5}$}    &  154, $10.78^{+0.01}_{-0.01}$, 0.35  &  298, $10.84^{+0.01}_{-0.01}$, 0.29    &  338, $11.11^{+0.01}_{-0.02}$, 0.34    &  926, $11.29^{+0.01}_{-0.02}$, 0.35      \\ 
\boldmath{$10^{10} < \frac{M_{*}}{M_{\odot}} < 10^{10.2}$}       &  188, $10.64^{+0.01}_{-0.01}$, 0.37  &  367, $10.68^{+0.02}_{-0.02}$, 0.33    &  494, $10.90^{+0.01}_{-0.02}$, 0.33    &  1018, $11.08^{+0.02}_{-0.01}$, 0.42    \\ 
\boldmath{$10^{9.5} < \frac{M_{*}}{M_{\odot}} < 10^{10}  \tnote{b}$}         &  691, $10.28^{+0.01}_{-0.01}$, 0.44  &  1095, $10.43^{+0.02}_{-0.02}$, 0.43  &  1561, $10.61^{+0.02}_{-0.03}$, $\la0.6$  &  3461, $10.69^{+0.02}_{-0.02}$, $\la0.7$    \\
\multicolumn{5}{c}{\textit{Quiescent Galaxies}} \\
\cmidrule(l{18em}r{18em}){1-5} 
\boldmath{$10^{11} < \frac{M_{*}}{M_{\odot}} < 10^{13}$}   &  89, $10.08^{+0.05}_{-0.04}$, $-$   &  138, $10.26^{+0.07}_{-0.06}$, $-$  &  114, $10.33^{+0.05}_{-0.05}$, $-$  &  255, $10.36^{+0.09}_{-0.06}$, $-$      \\ 
\boldmath{$10^{10} < \frac{M_{*}}{M_{\odot}} < 10^{11}$}   &  450, $10.00^{+0.03}_{-0.03}$, $-$   &  774, $9.85^{+0.09}_{-0.07}$, $-$  &  684, $9.96^{+0.10}_{-0.09}$, $-$    &  1591, $10.00^{+0.05}_{-0.08}$, $-$      \\
\bottomrule
\bottomrule
\end{tabular}
\begin{tablenotes}
\item [a] {\footnotesize Redshift bins are only shown up to $z\sim1$ where the majority of CO foreground comes from.}
\item [b] {\footnotesize The lowest-mass layers of faint star-forming and quiescent galaxies are not shown here as their mean and variance are less well-constrained.}
\end{tablenotes}
\end{threeparttable}
\end{center}
\caption{The number of galaxies, the mean total infrared luminosity and the scatter about the mean. The scatters of faint, star-forming galaxies in the two low-mass, high-redshift bins are shown as upper limits since in these cases the noise floor (both instrument and confusion) dominates the variance of the residual data cube. \label{tb:grand}}
\end{table*}
\endgroup

\begin{figure}[t!]
 \includegraphics[width=0.5\textwidth]{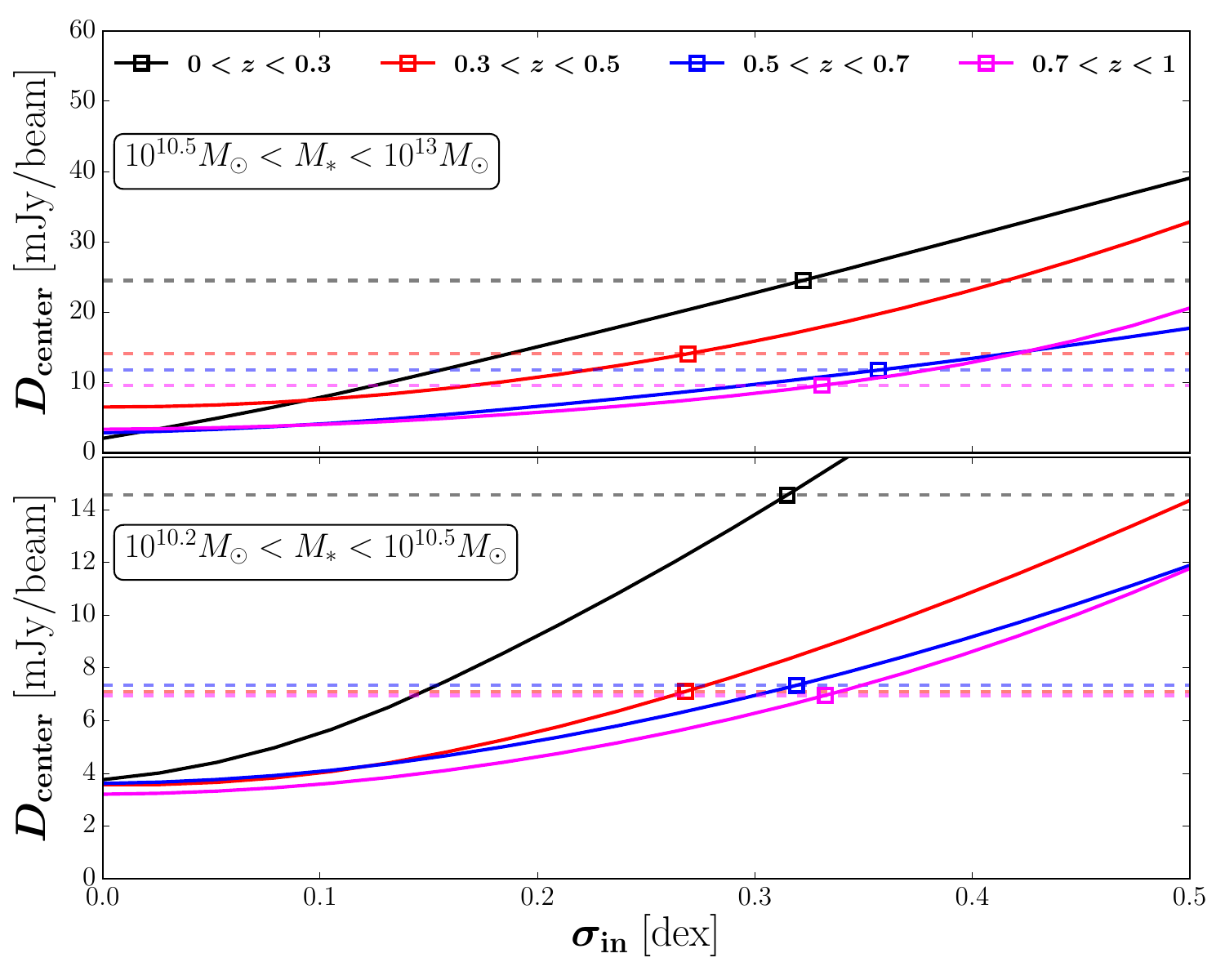}
 \caption{Calibration curves for the scatter in the derived $L_{\rm  IR}(M_{*},z)$ relation at 250\,$\mu$m. The top and bottom panels correspond to galaxies with stellar mass $10^{10.2}-10^{10.5}\,\rm M_{\odot}$ and $10^{10.5}-10^{13}\,\rm M_{\odot}$, respectively, in four redshift bins. 
The $x$-axis is the level of input scatter injected into the mock maps (labeled ``sigma'' in the panels of Figure~\ref{fig:map_illustration}), and the $y$-axis is the measured standard deviation level in the residual real/mock maps (illustrated by the color bar in Figure~\ref{fig:map_illustration}). The horizontal, dashed lines are the measured standard deviation levels in the residual real data cubes. The solid curves are the measured output standard deviation levels, with increasing input scatter, of the thumbnail-stacked residual cubes. The intersecting squares indicate the estimated level of scatter in the real sky images.}
 \label{fig:calib_cuve}
\end{figure}

Note that the layer of interest in the mock map is perturbed with different, adjustable levels of $\sigma_{S}$ (while all other layers are perturbed with the constant, fiducial value 0.3\,dex) to provide a ``calibration curve'' to compare with the real map. The idea is that the mock map is our best representation of the real map, including the positional source clustering and the scatter in luminosity that may be present in the actual galaxy populations, as well as instrument and confusion noise. From each of these, we want to subtract our best estimate of the average flux density in each layer, i.e. the \textit{unperturbed} {\footnotesize SIMSTACK} values. At this point, the residual real map will contain, at the positions of the sources in the layer of interest, information on the layer's intrinsic scatter in flux density. The magnitude of this scatter is then simply measured by gauging which level of $\sigma_{S}$ in the residual mock map matches the scatter in the residual real map. This is illustrated in Figure~\ref{fig:map_illustration}, where we show how the thumbnail-stacked mock data cube, $D_{\rm{mock}}^{k}$, compares to the real one, $D_{\rm{real}}^{k}$,  as we tune up the level of the scatter $\sigma_{S}$. For the purpose of measuring the scatter, we focus only on the central pixel of $D_{\rm{mock}}^{k}$ and $D_{\rm{real}}^{k}$. 

As shown in Figure~\ref{fig:calib_cuve}, the standard deviation in the mock thumbnail cubes gradually increases with increasing input scatter. The horizontal, dashed line represents the standard deviation measured in the real map. The scatter in the real maps can be consequently inferred from the intersection points, marked as squares in the figure. Since the maps are confusion-noise limited, the calibration curves do not start at zero, but rather at some noise floor equivalent to the standard deviation obtained by thumbnail stacking on random, non-source positions. A more detailed justification of this method based on end-to-end simulations is provided in Appendix~\ref{sec:A1}.

Table~\ref{tb:grand} lists the results of our extended {\footnotesize SIMSTACK} procedure, i.e. the number of galaxies in each bin, their mean total infrared luminosity and their scatter about the mean. In particular, we find an average logarithmic scatter, of $\langle \sigma_{\rm L} \rangle \sim 0.35$\,dex, with no evidence for systematic dependence on redshift or stellar mass, which is consistent with both observations \citep[e.g.][]{Whitaker_2012} and theoretical expectations \citep[e.g.][]{Kelson_2014,Sparre_2015} of the dispersion about the SFMS. \citet{Dutton_2010} investigate the origin of such small, roughly constant scatter in the SFMS using a semi-analytic model for disk galaxies based on smooth mass accretion onto dark matter halos and show that the scatter is mainly dominated by the variations in the gas accretion history and therefore does not evolve strongly with time or mass. Note that the method fails to give a reliable estimate of the scatter when the source population's flux density is too close to the noise floor.


\section{EVALUATING THE MASKING STRATEGY OF [C\,{\small II}] INTENSITY MAPPING EXPERIMENTS} \label{sec:3}

We will use the proposed configuration of TIME \citep{Crites_2014} as an example to demonstrate that the CO foreground can be efficiently removed by masking the contaminated voxels traced by infrared galaxies. Specifically, we apply our estimates of the mean and scatter in the $L_{\rm{IR}}(M_{*},z)$ relation to model CO emission in the $z<2$ sky to guide foreground masking. Based on our fiducial model, a robust detection of the [C\,{\small II}] signal can be achieved by masking galaxies using an evolving mass cut (roughly tracing a constant CO flux), which results in a moderate 4\%--8\% loss of the total survey volume. 

\subsection[]{Experiment Overview}

\begin{table}[t!]
  \footnotesize
  \centering
  \begin{tabular}{l|c}
    \toprule
    \multicolumn{2}{c}{\textbf{TIME Instrument Parameters}} \\
    \midrule
    Dish size & 12\,m \\
    Instantaneous FOV & $14' \times 0.43'$ \\
    Survey area & $1.3^{\circ} \times 0.43'$ (1$\times$180 beams) \\
    Number of spectrometers & 32: 16 per polarization \\
    Spectral range & 183--326\,GHz \\
    Spectral resolution & 90--120 \\
    Survey volume & 153\,Mpc$\times$1.1\,Mpc$\times$1240\,Mpc \\
    \bottomrule
  \end{tabular}
  \caption{TIME specifications. \label{tb:table_time}}
\end{table}

TIME is a high-throughput millimeter-wave imaging spectrometer array, designed to measure the 3D [C\,{\small II}] power spectrum. The clustering amplitude constrains the aggregate luminosity of [C\,{\small II}] emission from EoR galaxies. The instrument parameters of the proposed experiment are summarized in Table~\ref{tb:table_time}. 

\begin{figure}[h!]
 \includegraphics[width=0.5\textwidth]{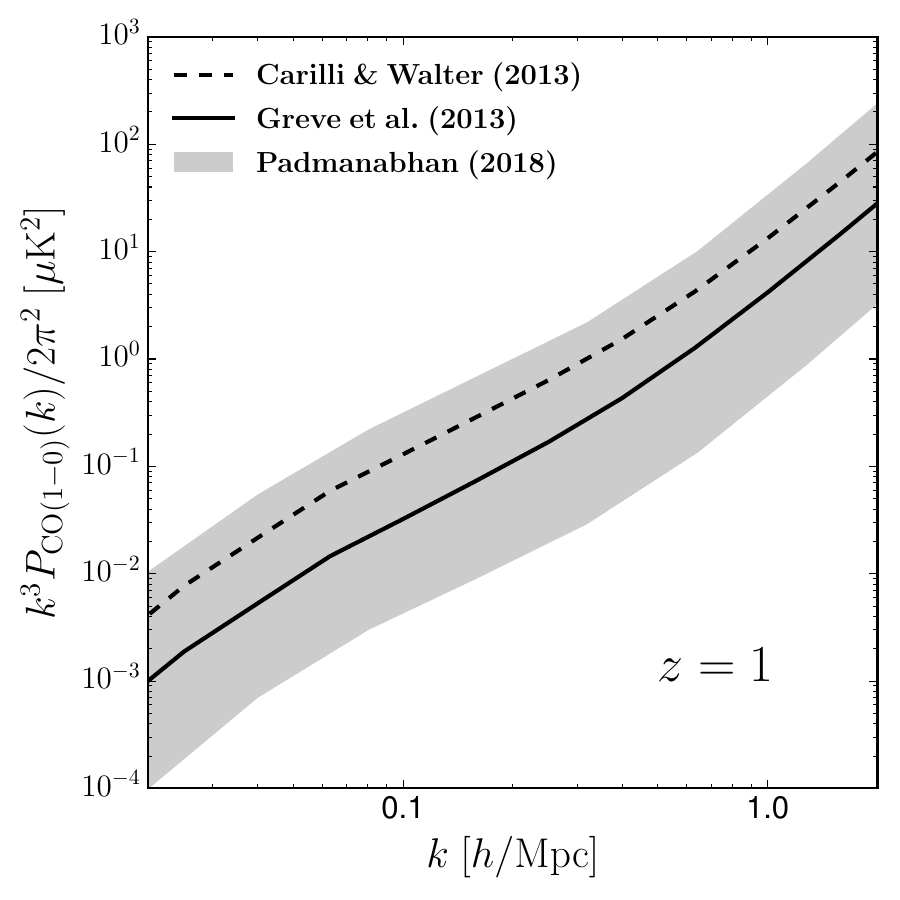}
 \caption{CO(1-0) power spectra predicted by our models assuming prescriptions of CW13 and G14 compared with that of the best-fit model calibrated to the observed CO luminosity function by \citet{Padmanabhan_2018}.}
\label{fig:CO_model}
\end{figure}

\subsection[]{Power Spectrum of CO Foreground}

\begin{figure*}[t!]
\centering
 \includegraphics[width=1.0\textwidth]{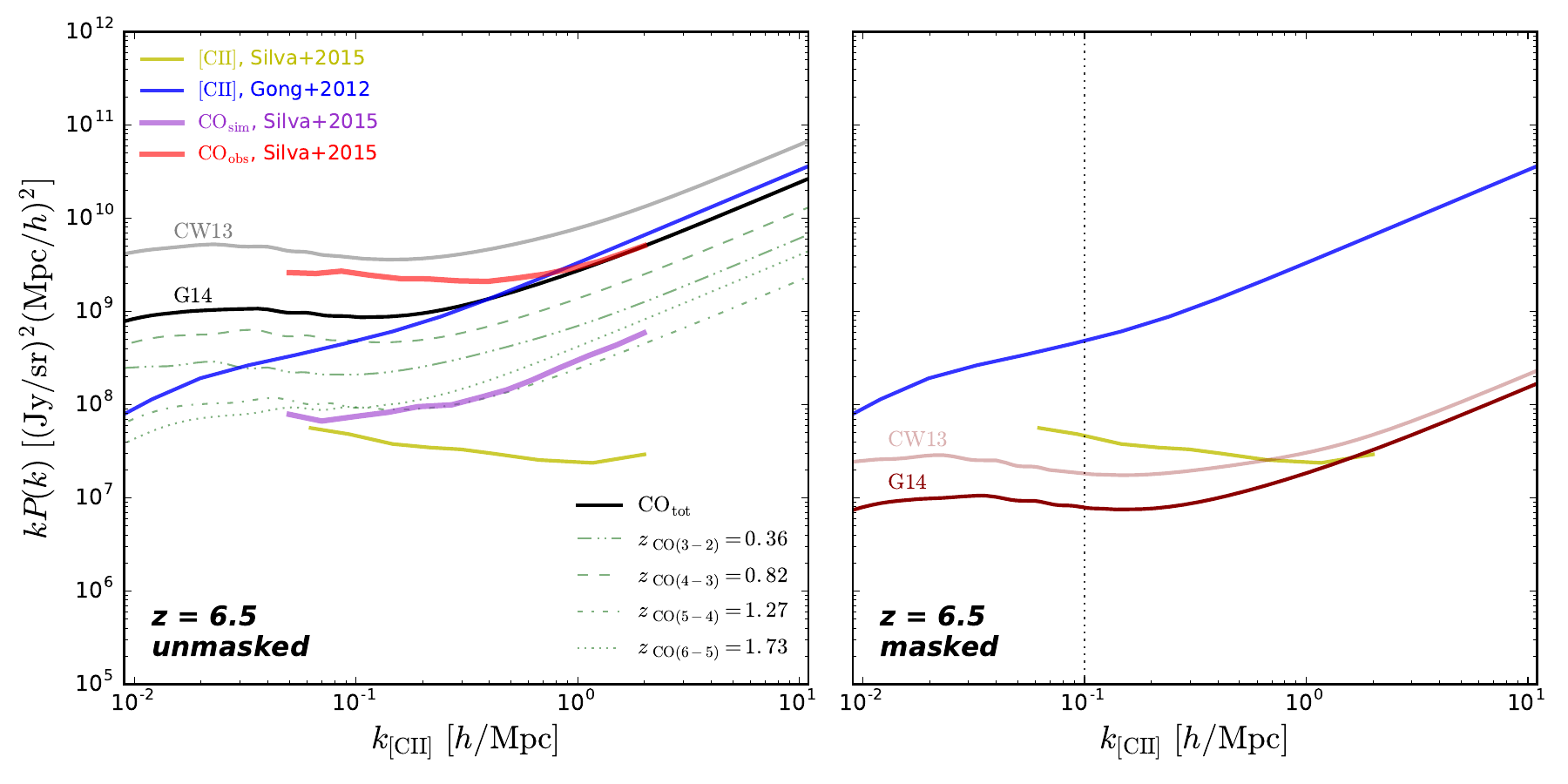}
 \caption{{\bf Left}: A comparison of the projected power spectra of unmasked CO emission lines and [C\,{\small II}] at $\nu_{\rm{obs}}\sim$250~GHz ($z_{[\mathrm{C}\,\text{\tiny{II}}]} \sim 6.5$).  
Our model illustrates the relative contribution to the total CO power spectrum (gray and black solid lines, respectively, for the two $L_{\rm{IR}}$--$L^{\prime}_{\rm{CO}}$ prescriptions, CW13 \& G14) from different CO transitions (green lines, just for G14), with a simulated input scatter of $\sigma_{\rm{tot}}=0.5$\,dex. Also shown are two CO models from \citet[][]{Silva_2015}: the first based on  simulated $L_{\rm{CO}}^{J}-M_{\rm{halo}}$ relations (purple line labeled \lq sim\rq); and the second from rescaling the observed infrared luminosity function (red line labeled \lq obs\rq). [C\,{\small II}] signals predicted by \citet{Gong_2012} and \citet[][``m2'' model]{Silva_2015}, shown as blue and yellow lines, respectively, highlight the large range of existing [C\,{\small II}] predictions. {\bf Right}: The same comparison as in the left panel, but after masking bright galaxies down to an evolving stellar mass cut (see Section~\ref{sec4p3}), which results in a $\sim$8\% (G14) and $\sim$18\% (CW13) loss of the total survey volume.
}
 \label{fig:mean_ps_z6p5}
 \end{figure*}

CO emission is derived for each object in the catalog by converting infrared luminosity to CO line strength with the well-established $L_{\rm{IR}}$--$L^{\prime}_{\rm{CO}}$ correlation (e.g., \citealt[hereafter CW13]{CW_2013}; \citealt[hereafter G14]{Greve_2014}; \citealt{MDZ_2015}). We can then use the measured stellar mass functions of galaxies at $0<z<2$ to calculate the power of CO line foregrounds, with the ability of monitoring different subsets (e.g., different stellar mass bins, quiescent vs star-forming galaxies, etc.) Now the total mean intensity of CO contamination can be expressed as
\begin{align}
\bar{I}_{\rm{CO}} = & \sum\limits_{J} \int_{M^{\rm min}_{*}}^{M^{\rm max}_{*}(z)} \mathrm d M_{*} \Phi(M_{*},z) \nonumber \\
& \times \frac{L_{\rm{CO}}^{J}\left(L_{\rm{IR}}(M_{*},z)\right)}{4 \pi D_{L}^{2}} y^{J}(z) D_{A}^{2}, 
\label{eq:mean_I}
\end{align}
where $M^{\rm min}_{*} = 1.0\times10^{8}\,\rm M_{\odot}$, $3 \leq J_{\rm upp} \leq 6$, representing all CO transitions acting as foregrounds and $\Phi(M_{*},z)$ being the stellar mass function measured by the COSMOS/UltraVISTA Survey \citep{Muzzin_2013}. $M^{\rm max}_{*}(z)$ represents an evolving mass cut that measures the depth of foreground masking (see Section 4.3 for a detailed discussion) and is set to $M^{\rm max}_{*,0}=1.0\times10^{13}\,\rm M_{\odot}$ when no masking is applied. The factor $y^{J}(z)=\mathrm{d}\chi/\mathrm{d}\nu_{\mathrm{obs}}^{J} = \lambda_{\mathrm{rf}}^{J}(1+z)^{2}/H(z)$  accounts for the mapping of frequency into distance along the line of sight \citep{VL_2010}. The comoving radial distance $\chi$, the comoving angular diameter distance $D_{A}$ and the luminosity distance $D_{L}$ are related by $\chi = D_{A} = D_{L}/(1+z)$. In the presence of scatter (as is always the case), the expectation value of a function $\mathcal{F}$ of CO luminosity at a best-fit $L_{\rm{IR}}(M_{*},z)$ given by {\footnotesize SIMSTACK} can be written as
\begin{equation}
E\left[\mathcal{F}\left(L_{\rm{CO}}^{J}\right)\right] = \int_{-\infty}^{\infty} \mathrm d x \mathcal{F}\left(10^{x}\right) P \left( x | \log L_{\rm{IR}} \right), 
\label{eq:lognormal_expectation}
\end{equation}
and
\begin{equation}
P \left( x | \log L_{\rm{IR}} \right) = \frac{1}{\sigma_{\rm{tot}} \sqrt{2 \pi}} \exp \left( - \frac{(x-\mu)^2}{2 \sigma_{\rm{tot}}^{2}} \right), 
\label{eq:lognormal}
\end{equation}
where
\begin{align}
\mu & = \log L_{\rm{CO}}^{J}(\log L_{\rm{IR}}) \nonumber \\ 
& = \alpha^{-1} (\log L_{\rm{IR}} - \beta) + \log r_{J} 
\end{align}
is derived from the best-fit $L^{\prime}_{\rm{CO}}$--$L_{\rm{IR}}$ correlation, with $r_{J}$ being some scaling factor for different $J$'s. Consequently, in the presence of scatter, Equation~\ref{eq:mean_I} becomes $\langle \bar{I}_{\rm CO} \rangle \equiv E\left[ \bar{I}_{\rm CO} (L_{\rm CO}^{J}) \right]$, which describes the expectation value of the total CO mean intensity, averaged over the probability distribution of $L_{\rm CO}^{J}$ as specified by $\mu$ and $\sigma_{\rm tot}$. In our calculation, we consider two prescriptions: (i) CW13, who give $\alpha=1.37\pm0.04$, $\beta=-1.74\pm0.40$, and scaling relations appropriate for sub-millimeter galaxies which are used to convert to transitions higher than $J=1 \rightarrow 0$, and (ii) G14, who provide $\alpha$ and $\beta$ coefficients for each individual $J$ transition (i.e., $r_{J}=1$) based on samples of low-$z$ ultra luminous infrared galaxies (ULIRGs) and high-$z$ dusty star-forming galaxies (DSFGs) comparable to CW13. Since the total CO foreground consists of multiple $J$ transitions, we deem the G14 prescription more appropriate for our purposes, because it treats both the slope and intercept as free parameters when fitting to galaxies observed in different $J$'s. Henceforth, we present our results based on the G14 model unless otherwise stated. 

It is worth noting that the $L^{\prime}_{\rm{CO}}$--$L_{\rm{IR}}$ relation is usually determined using a compilation of galaxy samples which collectively spans the stellar mass range $\log_{10}(M_{*})\sim$9.5-11.5. Simply extrapolating this relation to lower stellar masses without considering possible changes in the ISM in this regime likely overestimates the predicted CO emission, as observations suggest that local galaxies with $M_{*} < 10^9 M_{\odot}$ are deficient in CO, due to lower molecular gas contents and low metallicities (e.g., \citealt{Bothwell_2014}). \citet{MDZ_2015} find a 0.38\,dex scatter in $L_{\rm{IR}}$ for a given $L^{\prime}_{\rm{CO(1-0)}}$, which corresponds to 0.32\,dex scatter when converting infrared luminosity into CO luminosity. Comparable levels of scatter have also been identified by CW13 and G14 using galaxy samples of similar types. We note that different from our assumption in Equation~\ref{eq:lognormal}, \citet{Li_2016} re-normalize the log-normal distribution so that the \textit{linear} mean remains constant and that the level of scatter only affects the shot-noise component of the power spectrum. Instead, we choose to fix the \textit{logarithmic} mean in this work to best represent the distribution about the best-fit line for the observed $\log L_{\mathrm{IR}}$-$\log L^{\prime}_{\mathrm{CO}}$ correlation. Also, for simplicity, we ignore any potential correlation between the $\sim$0.3\,dex scatter intrinsic to the total infrared luminosity and the comparable $\sim$0.3\,dex scatter in the IR-to-CO conversion\,\footnote{This is a somewhat arbitrary choice given the potentially similar physics (star formation, dust attenuation, etc.) that leads to the observed scatters in both cases. As it is difficult to accurately determine this potential correlation, we simply assume here that 0.5\,dex is a relatively conservative estimate of the total scatter.} and combine them orthogonally (i.e. adding in quadrature, see also \citealt{Li_2016}), yielding a total scatter of $\sigma_{\rm{tot}} \sim 0.5$\,dex, which is what our reference model assumes hereafter. 

Figure~\ref{fig:CO_model} shows the CO(1$\mbox{--}$0) power spectra predicted by our CO model at $z=1$, compared with the best-fit model from \citet{Padmanabhan_2018} which is derived from abundance matching the halo mass function to the CO luminosity function observed at $0<z<3$. The overall power spectrum of the CO foreground can be written as the sum of the clustering and shot noise terms
\begin{equation}
P_{\rm{CO}}^{\rm{tot}}(z_{f},k_{f}) = P_{\rm{CO}}^{\rm{clust}}(z_{f},k_{f}) + P_{\rm{CO}}^{\rm{shot}}(z_{f},k_{f}), 
\end{equation}
where the clustering component can be derived from the mean intensity $I_{\rm{CO}}$, the average bias $\bar{b}_{\rm{CO}}(z)$ \citep{VL_2010} and the nonlinear matter power spectrum $P_{\delta \delta}^{\rm{nl}}$ (computed with the \texttt{CAMB}-based \texttt{HMFcalc} code, \citealt{Murray_2013}) as
\begin{equation}
P_{\rm{CO}}^{\rm{clust}}(z_{f},k_{f}) = \sum\limits_{J} \bar{b}^{2}_{\rm{CO}} \left(\bar{I}^{J}_{\rm{CO}}\right)^{2} P_{\delta \delta}^{\rm{nl}}(z_{f},k_{f}), 
\end{equation}
and the shot noise or Poisson component is given by
\begin{align}
P_{\rm{CO}}^{\rm{shot}}(z_{f}) = & \sum\limits_{J} \int_{M^{\rm min}_{*}}^{M^{\rm max}_{*}(z)} \mathrm d M_{*} \Phi(M_{*},z_{f}) \nonumber \\
& \times \left\{
\frac{L_{\rm{CO}}^{J}\left(L_{\rm{IR}}(M_{*},z_{f})\right)}{4 \pi D_{L}^{2}} y^{J}(z_{f}) D_{A}^{2} \right\}^2.
\label{eq:Pshot}
\end{align}

Estimating the CO contamination for any given observed [C\,{\small II}] power spectrum also requires rescaling (i.e. projecting) the corresponding CO comoving power spectrum at low redshift to the redshift of [C\,{\small II}]. Following \citet{VL_2010} and \citet{Gong_2014}, the projected CO power spectrum can be written as
\begin{equation}
P_{\rm{obs},CO}^{J}(z_{s},k_{s}) = P_{\rm{CO}}^{J}(z_{f},k_{f}) \times \left( \frac{\chi(z_{s})}{\chi(z_{f})} \right)^{2} \frac{y^{J}(z_{s})}{y^{J}(z_{f})},
\end{equation}
where $|\mathbf{k}_{f}| = \sqrt{( \chi(z_{s})/\chi(z_{f}))^{2}k_{\perp}^{2} + (y^{J}(z_{s})/y^{J}(z_{f}))^{2}k_{\parallel}^{2}}$ is the 3D k-vector at the redshift of CO foreground. Here we assume $k_{\perp} = \sqrt{k_{1}^{2}+k_{2}^{2}}$ and $k_{1}=k_{2}=k_{\parallel}$ for the 3D k-vector $|\mathbf{k}_{s}|=\sqrt{k_{\perp}^{2}+k_{\parallel}^{2}}$ at the redshift of [C\,{\small II}] signal. 

\begin{figure}[t!]
\centering
 \includegraphics[width=0.5\textwidth]{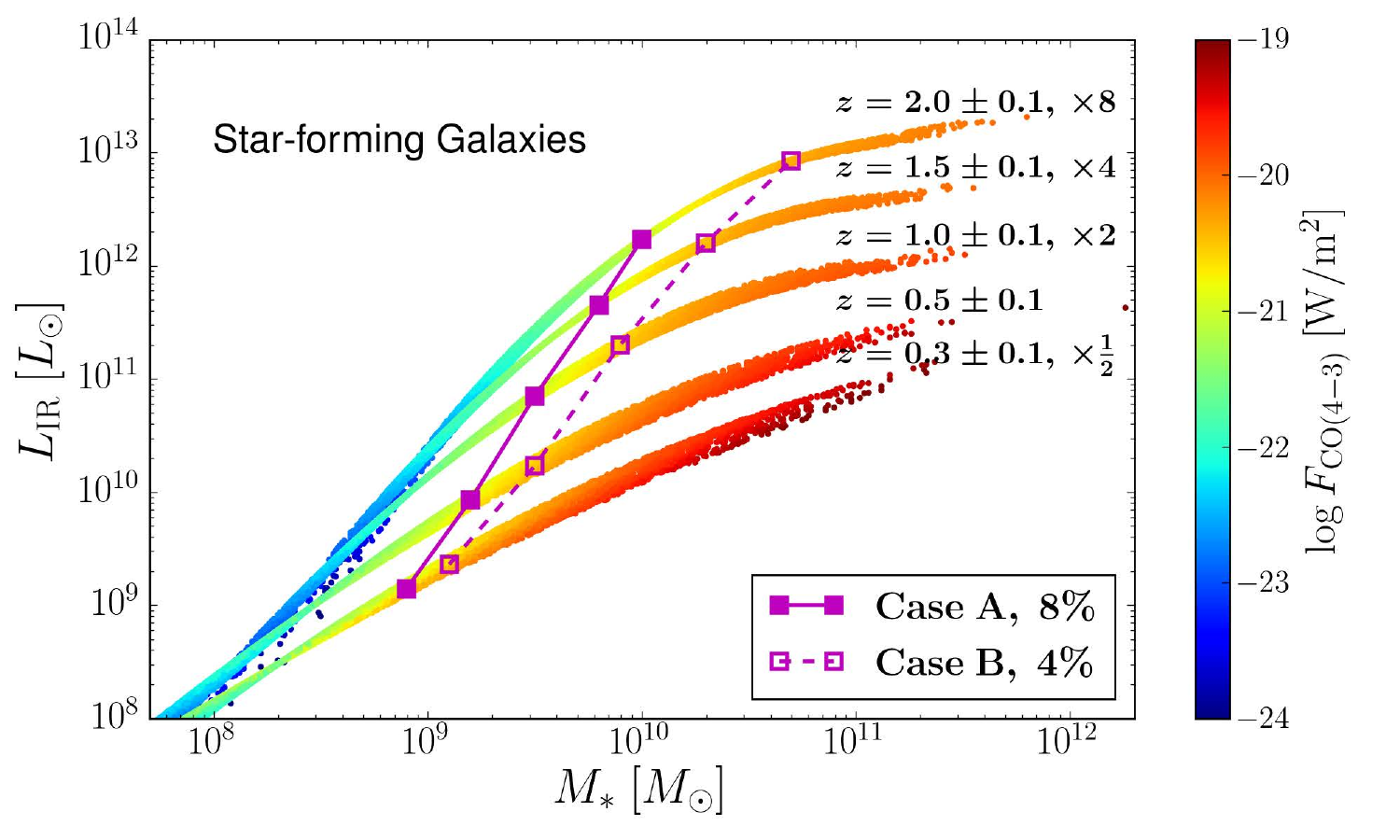}
 \caption{$L_{\rm{IR}}(M_{*},z)$ model predictions in five narrow redshift intervals, color-coded by the derived CO(4-3) flux in [W/m$^2$], assuming the G14 prescription. Each $(M_{*},z)$ point is taken directly from the UVISTA-DR2 catalog. Note that some bands are shifted vertically for visual clarity (the multiplicative factors are reported in the legend). The magenta curves are two examples of constant CO flux, or equivalently evolving stellar mass cut, corresponding to a total masked fraction of 8\% (Case A, extensive) and 4\% (Case B, moderate), respectively.}
 \label{fig:masking_strategy_1}
 \end{figure}

 \begin{figure}[t!]
\centering
 \includegraphics[width=0.45\textwidth]{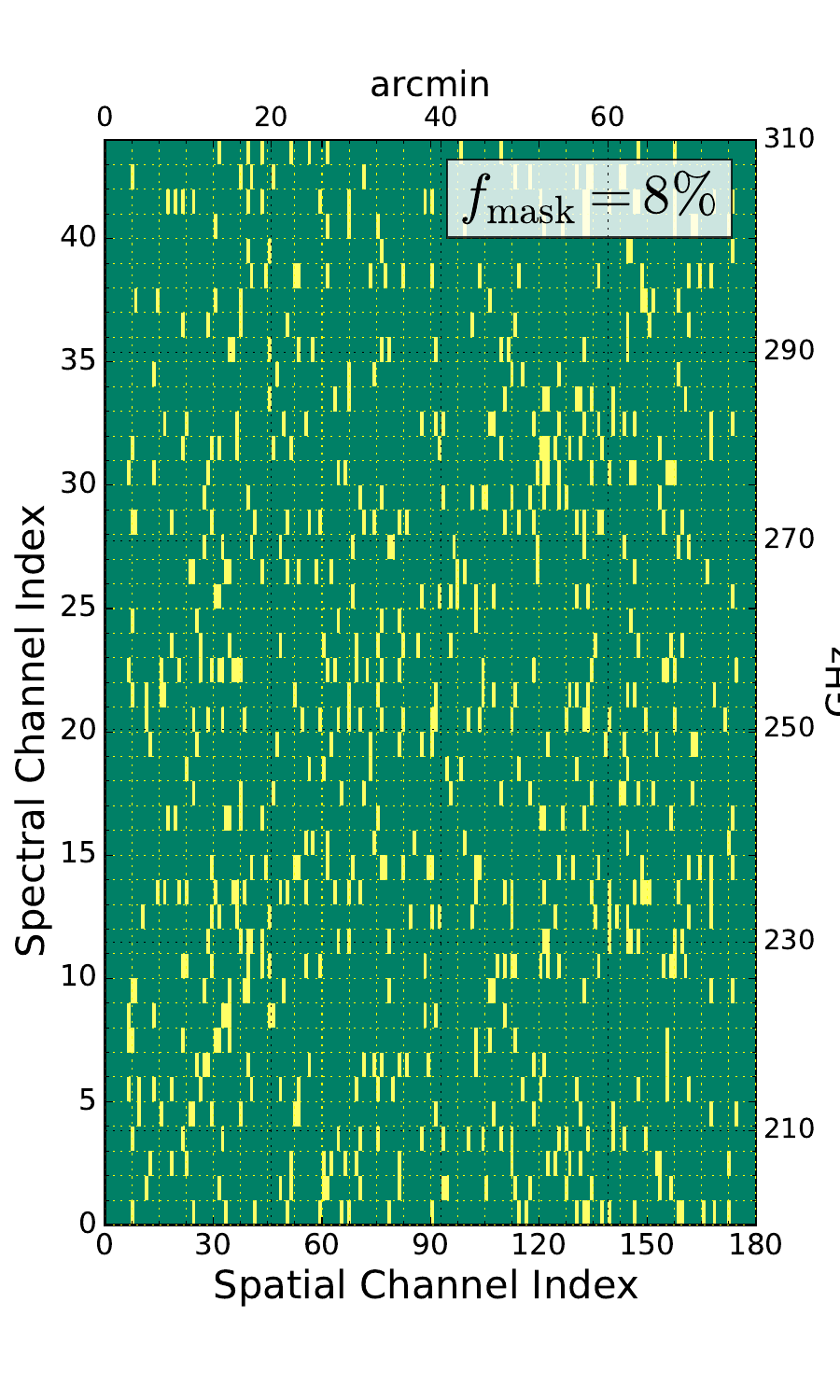}
 \caption{Voxel masking as a method of attenuating the CO foreground in [C\,{\small II}] intensity mapping experiments. By masking all the voxels that are contaminated by CO emission lines ($3 \leq J_{\rm upp} \leq 6$) from low-redshift galaxies with stellar mass higher than the evolving mass cut (two examples are shown in Figure~\ref{fig:masking_strategy_1}), we lose only a moderate fraction ($\la 8\%$) of our survey volume. The exact voxels being masked are illustrated in terms of their channel indices (44 spectral and 180 spatial channels) and are calculated from a mock TIME field chosen in the COSMOS/UltraVISTA field. Note that the spectral-to-spatial aspect ratio of the voxels here is set to 10 for visual clarity, while TIME's will be roughly 20.}
 \label{fig:Masking}
 \end{figure}

\begin{figure}[t!]
\centering
 \includegraphics[width=0.5\textwidth]{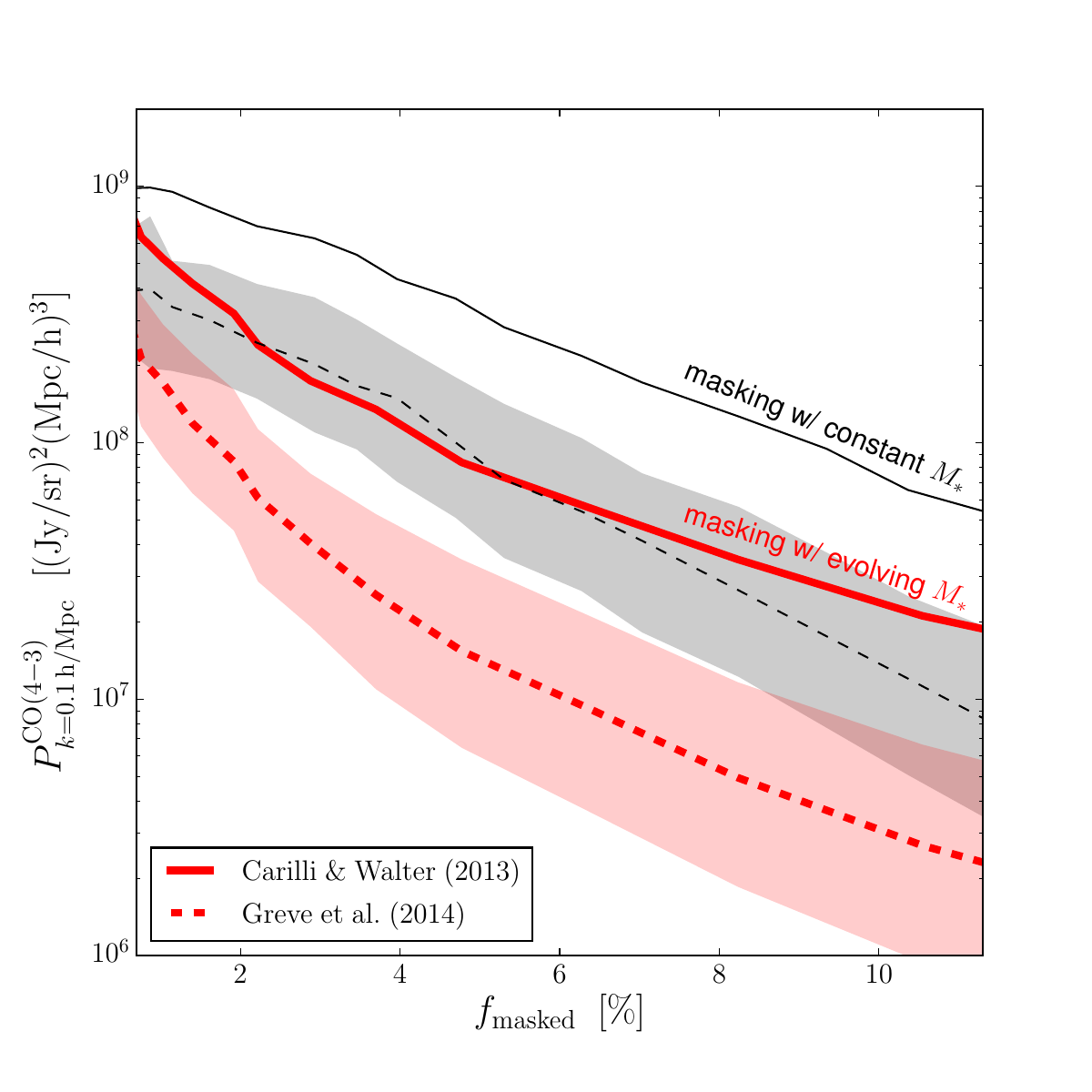}
 \caption{The predicted CO(4-3) power spectrum at $k=0.1$ $h$/Mpc (after scale-projecting into the frame of [C\,{\small II}] at $z\sim6.5$), as a function of voxel masking fraction for the two different masking strategies (constant, thin lines, vs. evolving $M_{*}$, thick lines; see text), and for the two different $L_{\rm{IR}}$--$L^{\prime}_{\rm{CO}}$ prescriptions considered in this work (CW13 and G14), showing that the evolving mass cut is more effective. The shaded bands represent the typical uncertainty in the inferred masking fraction due to fitting errors of the $L^{\prime}_{\rm{CO}}$--$L_{\rm{IR}}$ relation (only shown for G14 for clarity).}
 \label{fig:masking_strategy_2}
 \end{figure}

The left panel of Figure~\ref{fig:mean_ps_z6p5} shows our predicted CO power spectra projected into the frame of [C\,{\small II}] at redshift $z=6.5$ or an equivalent observing frequency of $\nu_{\rm{obs}}=250$ GHz. Contributions from different CO transitions (green curves) to the total CO power (gray and black solid curves) are shown by different line styles. For comparison, we also show two alternative CO models from \citet{Silva_2015}. The simulation-based model (purple line) is derived from fitting to the simulated $L_{\rm{CO}}^{J}-M_{\rm{halo}}$ relations \citep{Obreschkow_2009a,Obreschkow_2009b}, whereas the observational CO model (red line) is based on rescaling the observed infrared luminosity function \citep{Sargent_2012} with the ratios given by CW13. Finally, we note that \cite{Breysse_2015} assume ``Model A'' of \citet{Pullen_2013}, which models the CO luminosity at a given dark matter halo mass with a simple scaling relation and predict a much higher level of CO foreground for the [C\,{\small II}] signal given by the ``m2'' model of \citet{Silva_2015}. However, we note that the \citet{Pullen_2013} ``Model A'' is only optimized for observations at $z\sim2$ and fails to capture the transition to sub-linear scaling of the $L_{\rm{CO}}$--$M_{h}$ relation at halo masses $M_{h}>10^{11} M_{\odot}$.

The variation in modeling the [C\,{\small II}] intensity is illustrated by the predicted signals from \citet{Gong_2012} and \citet[][``m2'' model]{Silva_2015}, shown as blue and yellow lines, respectively. Recent ALMA observations of several typical star-forming galaxies at $5 \la z \la 8$ have tentatively suggested a high [C\,{\small II}]-to-IR luminosity ratio \citep{Capak_2015, Aravena_2016}. The [C\,{\small II}] luminosity function derived from these observations is similar to that of \citet{Gong_2012}, implying a high clustering amplitude. In terms of the cumulative number density of $z\sim6$ galaxies, the \citet{Gong_2012} model is also supported by recent observations \citep{Aravena_2016, Hayatsu_2017}, which suggest a cumulative number density more than 10 times higher for galaxies with $L_{[\mathrm{C}\,\text{\tiny{II}}]}>2\times10^{8}L_{\odot}$ compared to \citet{Silva_2015}. 

\subsection{Masking Strategy}\label{sec4p3}

3D positional information (RA, Dec, $z$) from the galaxy catalog allows us to remove spectral-spatial elements (voxels) in the survey. Namely, after a 3D intensity map consisting $\sim8000$ voxels has been measured, we discard the voxels contaminated by at least one foreground CO line falling into TIME's spectral range. We specifically use stellar mass as a measure of the masking depth because it is directly provided by the galaxy catalog and CO power spectra are conveniently parametrized in terms of it. This approach is different from the blind, bright-voxel masking approach \citep[e.g.][]{Breysse_2015}, which does not exploit spectral information to identify and mask the voxels contaminated by faint CO sources, and thus fails to reduce the CO foreground sufficiently.

Provided that the catalog is complete between the integration limits (i.e. $M^{\rm min}_{*} < M_{*} < M^{\rm max}_{*}$), it is possible to estimate the loss of survey volume at a given masking depth by simply counting the number of voxels contaminated by the CO lines emitted from galaxies to be masked. \citet{Laigle_2016} lists the 90\% completeness levels for the COSMOS/UltraVISTA (UltraDeep, or ``UD'') catalog under consideration here (also shown in Figure~\ref{fig:binning}); the stellar mass limits are $M^{90\%}_{*} \leq 10^{8.9}\, M_{\odot}$ for all CO transitions of interest. Since galaxies with $M_{*} \leq 10^{8.9}\, M_{\odot}$ contribute a negligible fraction ($\la 0.5\%$) of the total CO power, the loss fraction is essentially dominated by the choice of masking strategy. 

We optimize the masking sequence using an ``evolving mass'' cut, as shown in Figure~\ref{fig:masking_strategy_1}. Instead of masking galaxies with a simple, universal stellar-mass cut, which results in removing more voxels containing higher-redshift, relatively faint CO-emitters, in order to mask equally-massive, lower-redshift, CO-bright counterparts, we define a function $M_{*}^{\rm max}(z) \leq M^{\rm max}_{*,0}=1.0\times10^{13}M_{\odot}$  that is designed to follow a threshold of constant CO(4-3) flux (in W/m$^2$, assuming G14). Motivated by the range of uncertainty in [C\,{\small II}] models, we show two examples here corresponding to an extensive masking scheme (Case A) for the \citet{Silva_2015} model as well as a moderate masking scheme (Case B) for the \citet{Gong_2012} model. Masking essentially reduces the amplitude of CO power spectrum by varying the integration limit of the first and second CO-luminosity moments of the stellar mass function $\Phi(M_{*})$, which correspond to the mean intensity and shot-noise power, respectively. Namely, 
\begin{equation}
\frac{\langle \bar{I}_{\rm{CO}} \rangle_{\rm m}}{\langle \bar{I}_{\rm{CO}} \rangle_{\rm um}} = \frac{E[\int_{M_{*}^{\rm{min}}}^{M_{*}^{\rm{max}}(z)}L_{\rm{CO}}(M_{*})\Phi(M_{*})\mathrm d M_{*}]}{E[\int_{M_{*}^{\rm{min}}}^{M^{\rm max}_{*,0}}L_{\rm{CO}}(M_{*})\Phi(M_{*})\mathrm d M_{*}]}
\end{equation}
and
\begin{equation}
\frac{\langle P_{\rm{CO}}^{\rm{shot}} \rangle_{\rm m}}{\langle P_{\rm{CO}}^{\rm{shot}} \rangle_{\rm um}} = \frac{E[\int_{M_{*}^{\rm{min}}}^{M_{*}^{\rm{max}}(z)}L_{\rm{CO}}^{2}(M_{*})\Phi(M_{*})\mathrm d M_{*}]}{E[\int_{M_{*}^{\rm{min}}}^{M^{\rm max}_{*,0}}L_{\rm{CO}}^{2}(M_{*})\Phi(M_{*})\mathrm d M_{*}]},
\end{equation}
where the angle brackets indicate that values are averaged over the log-normal distribution described by Equations~\ref{eq:lognormal_expectation} and~\ref{eq:lognormal}.

\begin{figure*}[h!]
\centering
 \includegraphics[width=1.0\textwidth]{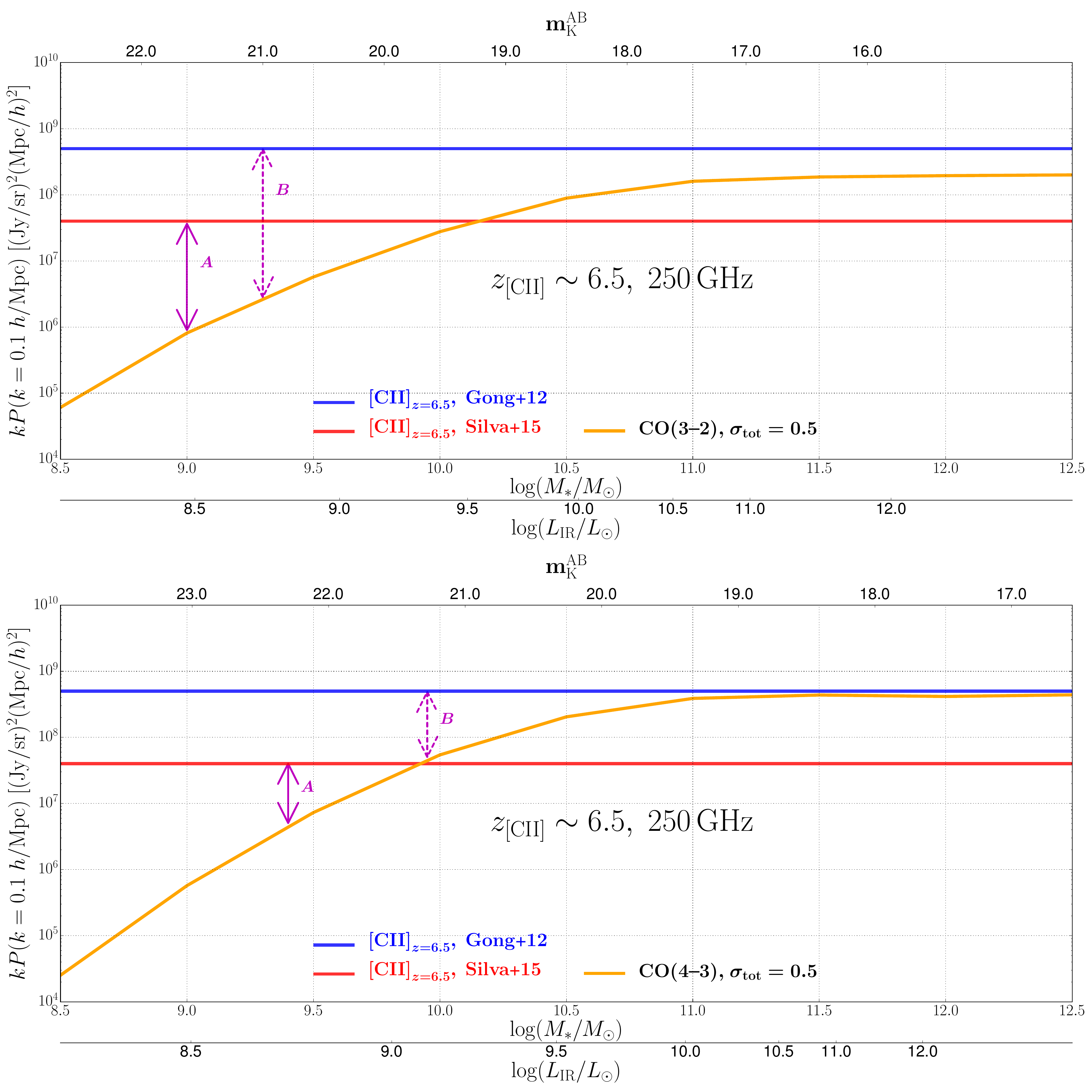}
 \caption{The predicted power at $k=0.1$ $h$/Mpc of [C\,{\small II}], CO(3-2) ({top panel}), and CO(4-3) ({bottom panel}), at redshift $z = 6.5$.
Multiple $x$-axes are shown to illustrate how the masking depth in K-band AB magnitude projects to $L_{\rm IR}$, stellar mass $M_{*}$, and mask fraction $f_{\rm mask}$. Note that the $\rm m^{\rm AB}_{\rm K}$ and $L_{\rm IR}$ scales differ from top to bottom panels because the interloping lines in the top and bottom panels originate from different redshifts: $z = 0.36$ and 0.82 for CO(3-2) and CO(4-3), respectively. Solid horizontal lines represent model predictions from \citet[][blue]{Gong_2012} and \citet[][``m2'', red]{Silva_2015}. The orange curve represents the CO power level vs.\@ masking fraction assuming a scatter of $\sigma_{\rm tot} = 0.5$\,dex. The solid (dashed) arrow indicates the evolving masking depth of Case A (Case B) considered in Figure~\ref{fig:masking_strategy_1}, which yields a [C\,{\small II}]-to-CO(3-2) power ratio of 50 (200) and a [C\,{\small II}]-to-CO(4-3) power ratio of 10 (10). 
}
\label{fig:moneyplot}
\end{figure*}
 
The constant vs. evolving stellar-mass cut approaches are explicitly compared in Figure~\ref{fig:masking_strategy_2}, where we show the predicted CO(4-3) power at $k=0.1$ $h$/Mpc (after scale-projecting into the frame of [C\,{\small II}] at $z\sim6.5$) for the two different masking strategies, and for the two different $L_{\rm{IR}}$--$L^{\prime}_{\rm{CO}}$ prescriptions considered in this work, namely CW13 and G14. One can see that there is a clear advantage in using the evolving mass cut strategy, yielding a CO contamination almost an order of magnitude lower than the constant mass cut (at equal masking fractions). We show this masking scheme for TIME voxels individually in Figure~\ref{fig:Masking}, where they are positioned according to their spatial ($x$-axis) and spectral channel ($y$-axis) indices. For the more extensive Case A masking, of which the depth decreases from $\sim 10^{9}M_{\odot}$ at $z\sim0.3$ to $\sim 10^{10}M_{\odot}$ at $z\sim2$, about 8\% of voxels need to be masked, as indicated by the yellow strips. 

In the right panel of Figure~\ref{fig:mean_ps_z6p5}, we show how this masking strategy can effectively bring down the CO contamination to levels that are sub-dominant to the clustering [C\,{\small II}] power. The power of total CO emission is calculated only from unmasked galaxies with an evolving stellar-mass cut, which results in a $\sim$8\% loss of the total survey volume (Case A). We note that although our analysis disfavors a total scatter larger than $\sim 0.5$\,dex, the uncertainty in the scaling relations of different rotational $J$ transitions among galaxy populations may also affect the predictions of CO power. Such uncertainty can be readily absorbed into the total scatter in our model by examining a broader range of scatter. 


The effect of voxel masking on the [C\,{\small II}] power spectrum is to essentially remove a small fraction of voxels from the survey volume in a nearly random (i.e. uncorrelated) pattern. In Appendix~\ref{sec:A2}, we demonstrate using a simulated light cone that simply discarding the CO-contaminated voxels would only cause a change in the raw, measured [C\,{\small II}] power spectrum of the order of the masked fraction ($\la 10\%$), which is already small compared to the expected measurement uncertainty and thus will not affect our predictions for the [CII]/CO power ratio.

In order to obtain the true power spectrum though, one must correct for the artifact arising from the coupling between Fourier modes due to windowing (i.e. masking) in real space \citep{Hivon_2002, Zemcov_2014}. Specifically, individual $k$ modes are propagated through the mask to characterize how their powers are mixed into other modes $k'$.  A mode-coupling matrix $M_{kk'}$ can be constructed consequently, whose inverse provides the appropriate transformation from a masked power spectrum to an unmasked one. Provided that mode mixing and other systematics such as instrument beam and experimental noise are properly corrected, the [C\,{\small II}] power spectrum should be measured in an unbiased way in the presence of voxel masking. Alternatively, the correlation information can also be extracted from the two-point correlation function (2PCF), which is formally the Fourier transform of the power spectrum. It has the advantage of being less affected by the complicated survey geometry and incomplete sky coverage due to masking, albeit making the theoretical interpretation less straightforward. A detailed discussion of such corrections and alternatives is beyond the scope of this paper and thus left for future work. 

We show in Figure~\ref{fig:moneyplot} the evolution of CO power at scale $k=0.1\,h/\rm{Mpc}$, where the clustering term dominates, with the masking depth expressed in K-band magnitude $m^{\rm AB}_{\rm K}$, infrared luminosity $L_{\rm{IR}}$ and stellar mass $M_{*}$. Two dominant CO transitions (3-2 and 4-3; see Figure~\ref{fig:mean_ps_z6p5}) are displayed separately here because the conversion between different masking depth expressions is redshift dependent. For our reference model, masking out voxels containing galaxies with $m^{\rm AB}_{\rm K} \la 22$ at $z<1$ renders a total CO power small enough compared with the [C\,{\small II}] clustering power with a moderate $\sim 8\%$ loss of total survey volume.

The accuracy of masking depends on the error in photometric redshift estimates with respect to instrument spectral resolution; for COSMOS DR2, $\sigma^{\rm{phot}}_{z}/(1+z^{\rm{phot}})$ is less than 1\% \citep{Laigle_2016}, comparable to TIME's typical voxel size in redshift space. While for simplicity the presented masked fractions are calculated assuming the maximum-likelihood photometric redshift, one may perform an even more conservative masking by accounting for the 68\% confidence interval of the photometric redshift distribution, which would approximately \textit{double} the masking fraction. Compared with the uncertainty in masking fraction due to fitting errors in the CO--IR relation shown in Figure~\ref{fig:masking_strategy_2}, photometric redshift errors would likely dominate the uncertainty in the predicted masking fraction.

As illustrated in Figure~\ref{fig:Masking}, we expect to be masking at most $\sim 700$ voxels at $0<z<2$ to reduce the level of CO contamination to a level required for a solid [C\,{\small II}] detection; hence, a follow-up campaign to measure spectroscopic redshifts is straightforward, if deemed necessary. For moderate masking (Case B; $\sim 350$ voxels), a typical $z\sim1$ star-forming galaxy close to our masking threshold $m^{\rm AB}_{\rm K} \sim 21$ requires about 3 hours of integration to obtain a robust spectroscopic redshift measurement with a multi-object spectrometer like MOSFIRE, which amounts to a total exposure time of about 30 hours for all $\sim 200$ galaxies\footnote{Note that the number of galaxies to follow up is lower than the number of voxels to be masked due to multiple CO transitions from the same source that fall within TIME's observing band.} that need to be masked within TIME's survey volume. For the more extensive masking (Case A; down to $m^{\rm AB}_{\rm K} \sim 22$), spectroscopic confirmation becomes more costly ($>60\ \mathrm{hours}$), so the masking of these fainter sources will be guided solely by photometric redshifts.

Finally, we note that this masking formalism is flexible enough so that it can be further optimized in multiple ways. First of all, stacking using more information of the sources (e.g., by including dust extinction, see Viero et al., in prep) than the mass-redshift plane could improve the total infrared luminosity model by reducing the scatter. Moreover, although here the masking depth is chosen quite arbitrarily to roughly trace a constant level of observed CO flux, it can be more formally optimized based on the properties of the foreground emitters, including the level of scatter. 

\subsection{Residual Foreground Tracers}\label{sec:residual}

Given the uncertainties in the strength of the [C\,{\small II}] signal and the CO contamination (see Figure~\ref{fig:mean_ps_z6p5}), it is desirable to probe the level of remaining CO foreground after the voxel masking technique is applied in order to determine whether the foreground has been removed sufficiently. \citet{Silva_2015} discuss the usefulness of cross correlation as a way to constrain the degree of post-masking foreground. Specifically, cross correlation can be done either between a foreground CO line and another dark matter tracer (e.g. a known population of galaxies) at the same redshift, or between two foreground CO lines (e.g. $J=4\rightarrow3$ and $J=3\rightarrow2$) emitted from the same redshift but contaminating the intensity maps observed at two different frequencies. The CO-galaxy cross-correlation requires an external dataset like COSMOS. The correlation can be checked as the masking depth increases. The CO--CO cross-correlation can be done within the experiment's own dataset, albeit at the expense of a potentially lower sensitivity after masking. The cross power in this case serves as a tracer of the degree of contamination as a function of masking depth. Since [C\,{\small II}] signals from different redshifts are uncorrelated, they do not contribute to the overall cross-correlation power. It is worth noting that these methods can test whether the CO foreground has been removed satisfactorily, although without indicating which sources must be further removed. In Appendix~\ref{sec:A3}, we present a more detailed discussion of the usefulness of cross-correlating CO lines from the same redshift, including how it can be used to measure CO lines themselves and thus constrain the cosmic molecular gas content.


\section{SUMMARY}
We presented a method to estimate the mean and scatter of CO line emission from measurements of the total infrared luminosity, $L_{\rm{IR}}$, and showed how it can be applied as a foreground removal strategy for [C\,{\small II}] intensity mapping experiments. We optimized the trade-off between the relative strength of CO/[C\,{\small II}] power and the loss of survey volume. We found that even in the most conservative scenario, by progressively masking galaxies above a stellar mass cut increasing with redshift --- which approximately amounts to K-band magnitudes of $m_{\rm{AB}} \la 22$ at $z < 1$, or $\sim$8\% of all voxels --- a [C\,{\small II}]/CO power ratio $\ga 10$ is achievable in the clustering amplitude. 

\acknowledgments
The authors would like to thank the anonymous referee for valuable suggestions. The authors also acknowledge Ryan Quadri and Adam Muzzin for the continued support of our stacking program and the valuable insights from the near-infrared community. T.-C. C. acknowledges MoST grant 103-2112-M-001-002-MY3 and JPL R\&TD Award 01STCR - R.17.226.063. JH is supported by the National Science Foundation Graduate Research Fellowship under Grant No. DGE-1144469. AC was supported by a KISS postdoctoral fellowship and is now supported by the National Science Foundation Astronomy and Astrophysics Postdoctoral Fellowship under Grant No. 1602677. MBS acknowledges the Netherlands Foundation for Scientific Research support through the VICI grant 639.043.006. MPV acknowledges support by the US Department of Energy through a KIPAC Fellowship at Stanford University. 

\newpage

\bibliographystyle{bibapj}
\bibliography{masking}

\appendix

\section{Validation of the stacking method}\label{sec:A1}

In order to demonstrate the validity of our methods for estimating the mean and scatter of source populations, and to identify any potential bias due to galaxy clustering, we apply the procedures from the previous section to simulated maps, whose input mean flux densities and their scatter are known. We base our simulations on the COSMOS catalog which inherently contains the positional information about source clustering and defer a more thorough analysis involving a varying degree of clustering to future work.

Simulated FIR/sub-mm maps are generated in the same way as the mock maps for scatter characterization, as described in Section~\ref{calibration_form}. We note that different realizations of random flux assignments ensure that we obtain distinct maps with similar statistics. Flux densities are assigned to sources in each individual $(M_{*},z)$ bin according to a log-normal distribution, with scatter $\sigma_{\rm{in}}$ and mean $\langle S_{\rm{in}} \rangle$. 

\begin{figure}[t!]
 \centering
 \includegraphics[width=0.5\textwidth]{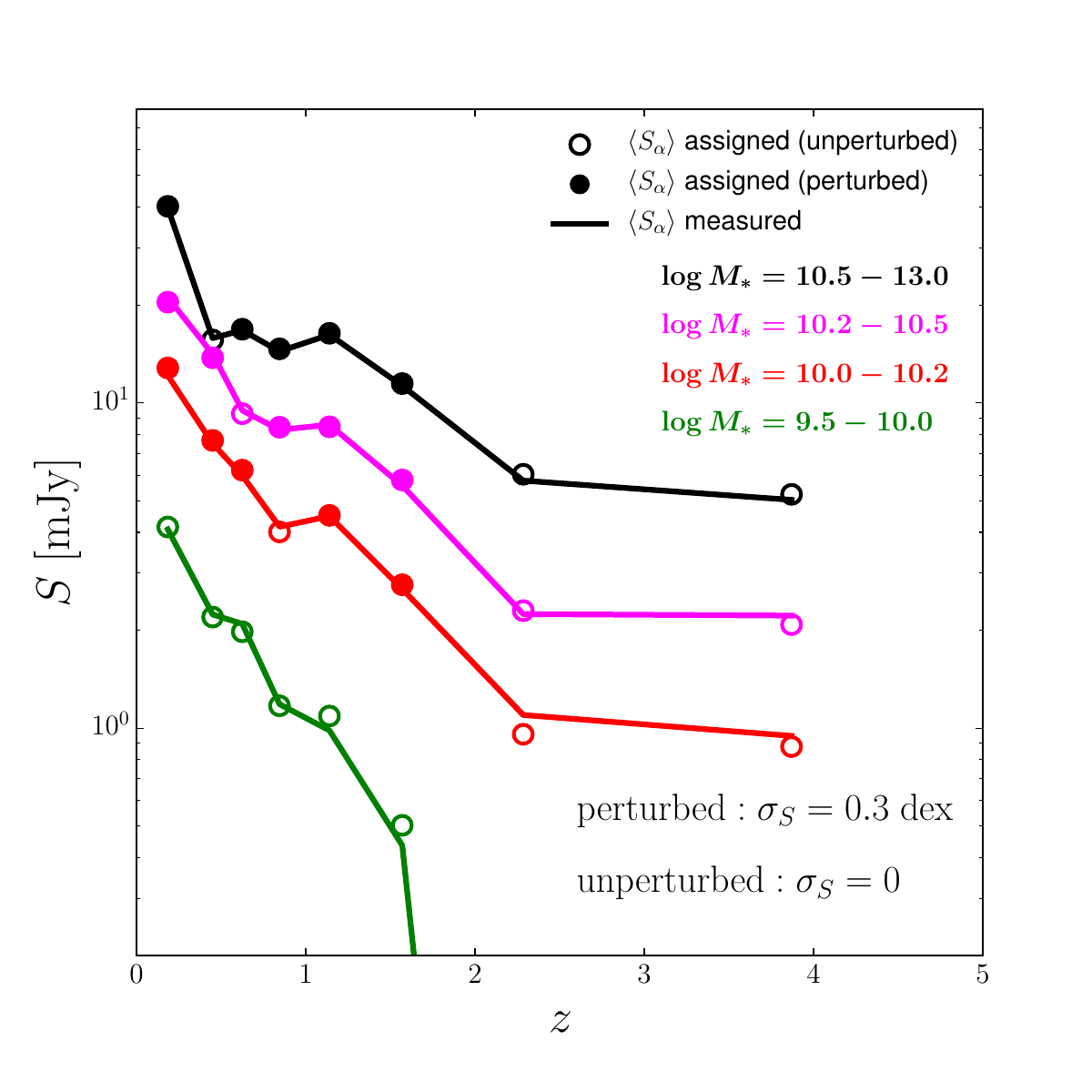}
 \caption{Robustness of measuring the mean flux densities with {\footnotesize SIMSTACK} at 250\,$\mu$m. When a simulated map is created, sources in a selected number of bins are assigned flux densities according to a log-normal distribution with a fiducial scatter of 0.3\,dex (i.e. sources are perturbed by 0.3\,dex of scatter), whose means are calculated and represented by the filled circles. Sources in other bins are simply assigned a fixed flux density with zero scatter (i.e. sources are unperturbed) and their means are represented by the open circles. {\footnotesize SIMSTACK} measurements of the constructed simulated map shown by the solid lines are then compared with both the filled and open circles, indicating a good agreement between the assigned fluxes (both perturbed and unperturbed) and the measured ones at all redshifts and flux levels. }
 \label{fig:stack_sim}
\end{figure}

\begin{figure}[t!]
 \centering
 \includegraphics[width=0.5\textwidth]{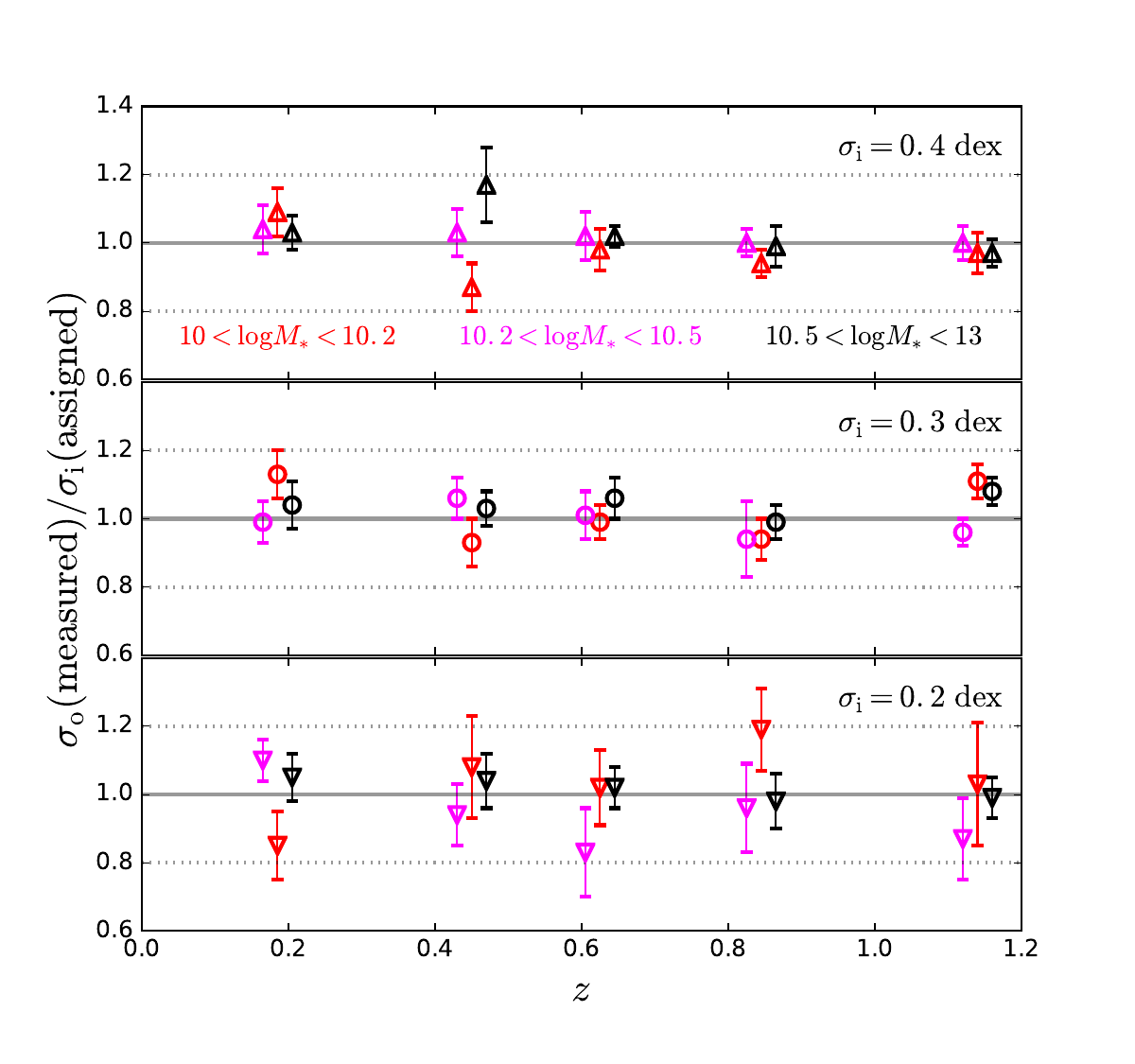}
 \caption{Comparison between the measured and assigned levels of scatter as a validation of the scatter measurements (i.e. 1 stands for a perfect agreement). The middle panel shows the ratio of the scatter measured by our thumbnail stacking formalism to that assumed by the distribution of fluxes (0.3\,dex) to generate simulated maps. This ratio of recovered to assigned scatters ($\sigma_{\rm{o}}/\sigma_{\rm{i}}$) is evaluated for different stellar mass and redshift bins, as indicated by the $x$-axis and marker colors. For comparison, the top and bottom panels show the cases where, only in the bin under examination, the fluxes assigned to the galaxies are drawn from a log-normal distribution with a different scatter (0.4 and 0.2\,dex, respectively). Sources not in the bin under examination are varied by 0.3\,dex in all three cases. Note that the data points are slightly shifted along the $x$-axis for visual clarity.}
\label{fig:scatter_sim}
\end{figure}

We further verify that {\footnotesize SIMSTACK} estimates are robust for subsets with different levels of scatter. As shown in Figure~\ref{fig:stack_sim}, flux densities $\langle S_{\rm{out}} \rangle$ estimated with or without a 0.3\,dex scatter (solid line) are consistent with their respective mean inputs (circles) for a simulated map. Note that in these simulations we only perturb a subset of bins (indicated by the filled circles), in order to test the interplay between perturbed and unperturbed layers in the mock maps. Additionally, we test that {\footnotesize SIMSTACK} estimates are unbiased for perturbations of up to 0.5\,dex on different combinations of bins (not shown), which provides confidence that the mean flux density distribution in the actual $\sim 0.35$\,dex measurement is correctly estimated. 

As a final validation test, we estimate the scatter in simulated maps. Figure~\ref{fig:scatter_sim} shows the ratio of measured scatter ($\sigma_{\rm{o}}$) to the assigned scatter ($\sigma_{\rm{i}}$), for different stellar mass/redshift bins. The error bars indicate the 68\% confidence intervals estimated from many map realizations. We investigate the robustness of this method with two simple tests: (i) a given bin is assigned a scatter different from the fiducial value (0.3\,dex), but still within the range (0.2--0.4\,dex) observed, and (ii) a different fiducial scatter within the observed range is assigned to the ``background'' sources. The first test is shown in the top and bottom panels of Figure~\ref{fig:scatter_sim}, where the bins under examination are perturbed by $\sigma_{\rm{i}} = 0.2$ and 0.4\,dex respectively, along with the case shown in the middle panel where all bins have the same 0.3\,dex scatter. Our method recovers the input scatter to within $\sim$\,10\%, typically, and 20\% at worst. For the second case, we find that, on average, varying the fiducial scatter between 0.2 and 0.4\,dex introduces less than 10\% uncertainty in the recovered scatter, comparable to the level of statistical error. Therefore, although a fiducial scatter must be assigned to properly account for the flux variance due to ``background'' sources in a confusion-limited map, our method is generally insensitive to its exact value, at least within the range of the observed scatter in the SFMS. 

\section{Effect of masking on the [C\,{\small II}] power spectra}\label{sec:A2}

Intensity maps of [C\,{\small II}] emission from the EoR will be contaminated by emission from several CO transitions at low redshifts whose signal is expected to be higher than that of the target [C\,{\small II}] line. Masking voxels contaminated by strong CO emission has been shown to significantly reduce the foreground lines signal. 

During the CO masking process, a fraction of the [C\,{\small II}] signal will be inevitably removed. Given that CO and [C\,{\small II}] emissions are originated from different volumes in space, they will be observed as uncorrelated both in angular position and in the observed frequency. Therefore, the percentage of reduction of the [C\,{\small II}] intensity due to the masking procedure should be of the order of the percentage of pixels masked, while the CO intensity of emission will be substantially reduced as long as the bright CO galaxies are correctly identified. The masked pixels can also be seen as a loss in volume of the observed field and the [C\,{\small II}] corrected for masking such as is done in CMB studies. This correction will be done for observational data allowing for the recovery of the target signal as long as the masked percentage is not very high. For this study we are however not going to discuss the possible algorithms that can be used to correct for this masking since even without the correction the target signal would be reasonably well recovered for the discussed masking percentages. 

We simulate the masking procedure using a CO signal characterized by the \cite{Greve_2014} model and for two models of [C\,{\small II}] emission. The CO and [C\,{\small II}] lines are then masked according to a cut in stellar mass corresponding to the Case A masking described in this paper. This corresponds to a masking of about $10\%$ of the simulated volume. 

The line signals are obtained by post processing galaxy data from the EAGLE simulation \citep{Schaye_15, McAlpine_16, Crain_17} using semi-analytic models. The stellar masses predicted by this simulation differ from that of the COSMOS/UltraVISTA survey as shown in Figure 1 from \cite{Furlong_15}. However, the qualitative conclusions that can be taken from this exercise are valid anyway. 

\begin{figure}[t!]
\begin{centering}  
\includegraphics[angle=0,width=0.5\textwidth]{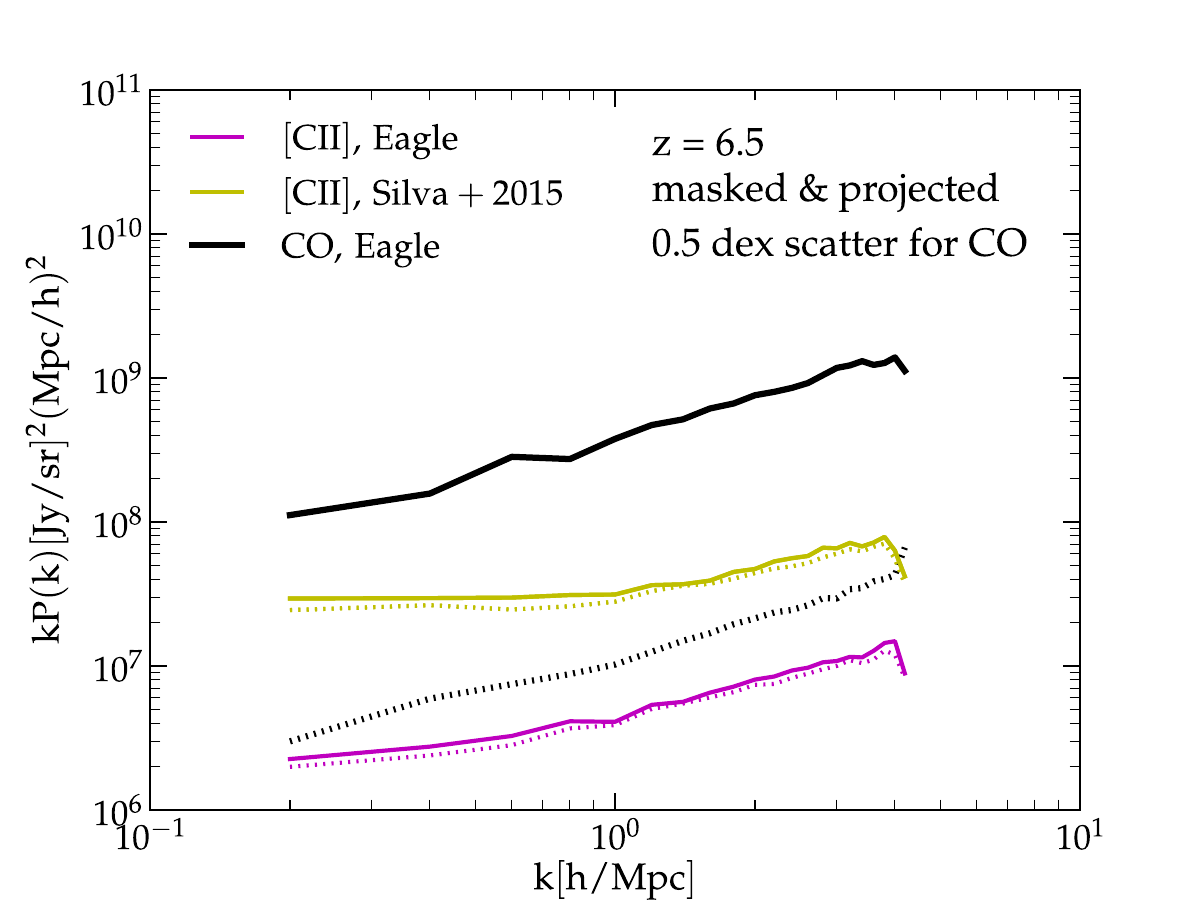}
\caption{Power spectra of CO (projected) and [C\,{\small II}] emission computed from simulated intensity maps before (solid) and after (dotted) Case A masking as illustrated in Figure~\ref{fig:masking_strategy_1}.}
\label{fig:COCIImasked}
\end{centering}
\end{figure}

The IR luminosities of CO emitting galaxies are obtained using Equation~\ref{eq:lmz_e1} (where passive galaxies were identified as galaxies with $\dot{M}_{*} = 0\,\mathrm{M_{\odot}yr^{-1}}$). Note that given the resolution of the EAGLE simulation, the star formation rate (SFR) is only resolved for  $\dot{M}_{*} > 2 \times 10^{-3}\,\mathrm{M_{\odot}yr^{-1}}$. Therefore, in this simulation some galaxies might have been considered as having a quenched SFR while still having some star formation. The CO luminosities are then derived from the IR luminosities using the relations by \cite{Greve_2014} and assuming a total scatter of 0.5\,dex in this relation. 

We model the [C\,{\small II}] luminosities assuming the following relation:
\begin{equation}
L_{\rm [C\,{\small II}]}\,[L_{\odot}] = 9.22 \times 10^6 \dot{M}_{*}\,[{\rm M_{\odot} yr^{-1}}].
\end{equation}
The [C\,{\small II}] signal is calculated assuming the relation between SFR and halo mass from Silva et al 2015 (where the halo masses were taken from the EAGLE simulation) or directly assuming the SFRs from the Eagle simulation. These two models span the expected uncertainty on the [C\,{\small II}] signal during the EoR (more precisely at $z=6.5$) due to the uncertainty on the SFR powering these emissions. Another important source of uncertainty on the amplitude of the [C\,{\small II}] signal is the evolution of the ratio between IR luminosity and [C\,{\small II}] luminosity towards high redshifts, which is however beyond the scope of this paper. 

Figure \ref{fig:COCIImasked} shows the effect of masking pixels on the CO and on the CII power spectra. According to these CO/[C\,{\small II}] models, the masking described in this paper would reduce the CO signal efficiently. The relative amplitude of the masked CII signal to the CO signal will mainly depend on the initial relation between the amplitude of the two signals. 

\section{Cross correlating [C\,{\small II}]+CO maps}\label{sec:A3}

As mentioned in Section~\ref{sec:residual}, the cross correlation between maps of [C\,{\small II}]+CO emission can be used to test if the masking procedure effectively decreased the signal of some of the line contaminants. Moreover, without masking, this cross correlation can be used to get an independent measurement of the intervening CO lines themselves.

In the frequency range covered by TIME surveys there are a few sets of two observing frequencies which contain emission from two or more adjacent CO lines originating from the same redshift. As an example the [C\,{\small II}] intensity maps at $z=7.8$ and 5.6 will be respectively contaminated by CO(3-2) and CO(4-3) lines emitted from $z \sim 0.6$. Since only two lines emitted from the same redshift will be correlated, this cross correlation in principle only measures the CO foreground. 

\begin{figure}[t!]
\begin{centering}  
\includegraphics[angle=0,width=0.5\textwidth]{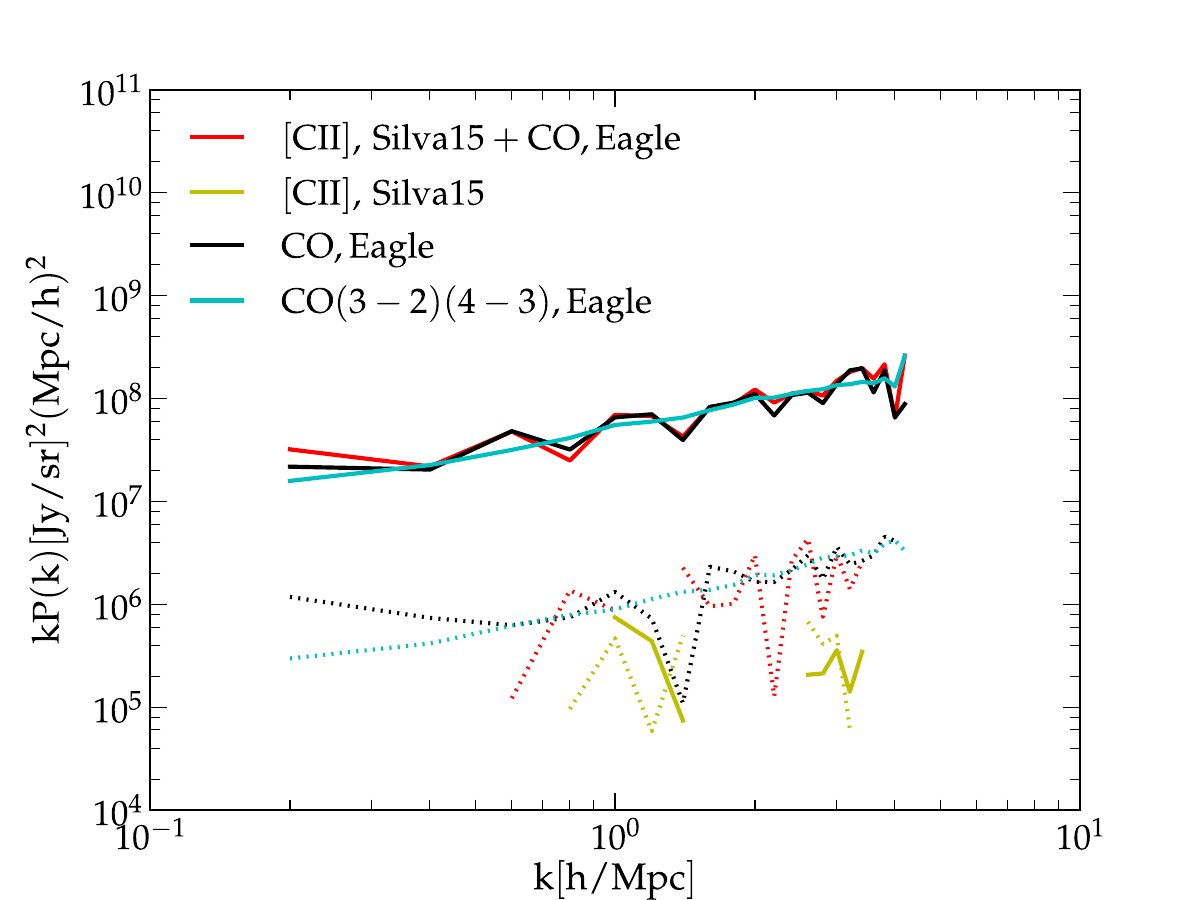}
\caption{Cross correlation power spectra between observed intensity maps at frequencies 216.1\,GHz ($z_{\rm [C\,{\small II}]} = 7.8$) and 288.2\,GHz ($z_{\rm [C\,{\small II}]} = 5.6$), corresponding to the observed frequencies of CO(3-2) and CO(4-3) lines emitted from $z\sim0.6$. The solid and dashed lines represent power spectra before and after masking respectively. Different colors indicate cases where the simulated intensity maps contain different combinations of signal and foreground lines.}
\label{fig:cross}
\end{centering}
\end{figure}

In terms of a tracer of residual CO emission, the amplitude of the cross correlation of the two \textit{masked} signals will be proportional to the product of the residual signals from the two CO lines. The shape of the cross correlation power spectra, between the two masked signals, will be correlated and uncorrelated at different scales if masking has reduced the CO foreground sufficiently. This lack of correlation is a strong indication that the masked maps are dominated by the [C\,{\small II}] emission. In this case, the nonzero power is due to the self correlations of the emission within individual simulation boxes, which can be understood as high-order terms in the cross correlation. Figure \ref{fig:cross} shows this cross correlation power spectra made with the simulations described in Section \ref{sec:A1}. 

On the other hand, the cross correlation of the two \textit{unmasked} signals will result in the product of the signals from the two CO lines and serve as a probe of CO intensities, which can also be further converted into $\rm H_2$ mass to infer the molecular gas content of galaxies. It should be noted, however, that certain assumptions of CO excitation have to be made in order to understand the correlation factors (i.e. line ratios) of different CO transitions and therefore interpret the cross correlation measurements of adjacent CO lines. Fortunately, existing observations suggest rather small variations in the line ratios of adjacent CO lines \citep[e.g.,][]{CW_2013}, allowing [C\,{\small II}] experiments like TIME to make reliable measurements of CO lines by cross-correlating within the dataset.

\end{document}